\documentclass[12pt,halfline,a4paper]{ouparticle}
\usepackage[round]{natbib}
\usepackage{hyperref}

\newenvironment{itemize*}%
  {\begin{itemize}%
    \setlength{\itemsep}{0pt}%
    \setlength{\parskip}{0pt}}%
  {\end{itemize}}

\newenvironment{enumerate*}%
  {\begin{enumerate}%
    \setlength{\itemsep}{0pt}%
    \setlength{\parskip}{0pt}}%
  {\end{enumerate}}

\newcommand{\E}{\mathbb{E}}

\newcommand{\thetahat}{\hat{\theta}}

\newcommand{\ind}{\mathbf{1}}

\DeclareMathOperator*{\argmax}{arg\,max}

\newcommand{\gammaswf}{\gamma\operatorname{-SWF}}

\newcommand{\betaswf}{\beta\operatorname{-SWF}}

\newcommand{\qbar}{\overline{q}}

\newcommand{\qubar}{\underline{q}}

\newcommand{\R}{\mathbb{R}}

\newcommand{\C}{\mathcal{C}}

\newcommand{\Q}{\mathcal{Q}}
\newcommand{\SWF}{\operatorname{SWF}}

\newtheorem{theorem}{Theorem}

\newtheorem{corollary}{Corollary}
\newtheorem{result}{Result}

\newtheorem{definition}{Definition}
\newtheorem{example}[definition]{Example}

\newtheorem{proposition}{Proposition}

\newtheorem{assumption}[theorem]{Assumption}

\begin{document}


\title{\Large \textsc{Optimal Defaults, Limited Enforcement \\ and the Regulation of Contracts}\footnote{We thank Oliver Hart and Eric Maskin for their invaluable feedback. We also thank Ed Glaeser, Ben Golub, Jerry Green, Elhanan Helpman, Shengwu Li, Matthew Rabin, Andrei Shleifer and Anthony Lee Zhang for insightful comments.}}

\author{Zo\"{e} Hitzig and Benjamin Niswonger\footnote{Hitzig: Department of Economics, Harvard University, 1805 Cambridge Street, Cambridge, MA 02138 (e-mail: zhitzig@g.harvard.edu). Niswonger: Department of Economics, Harvard University, 1805 Cambridge Street, Cambridge, MA 02138 (e-mail: niswonger@g.harvard.edu).}}



\abstract{We study how governments promote social welfare through the design
of contracting environments. We model the regulation of contracting as \textit{default delegation}: the government chooses a \emph{delegation set} of contract terms it is willing to enforce, and influences the \emph{default} terms that serve as outside options in parties' negotiations. Our analysis shows that limiting the delegation set principally mitigates externalities, while default terms primarily achieve distributional objectives. Applying our model to the regulation of labor contracts, we derive comparative statics on the optimal default delegation policy. As equity concerns or externalities increase, \textit{in-kind} support for workers increases (e.g. through benefits requirements and public health insurance). Meanwhile, when worker bargaining power decreases away from parity, support for workers increases \textit{in cash} (e.g. through cash transfers and minimum wage laws).



\vspace{3mm}

\noindent \textbf{JEL:} D82, D86, D63}


\date{\today}


\maketitle
\thispagestyle{empty}
\vfill

\pagebreak

\setcounter{page}{1}
\section{Introduction}

Contracting parties do not negotiate in a vacuum. Their negotiations play out in specific legal and social environments which explicitly and implicitly bound the range of contracts that can be written. 

Constraints on contracting take two primary forms. First, the law limits the set of enforceable contracts. If a contract term violates a so-called ``immutable rule" in contract law, it will not be enforced in court. For instance, in the United States, courts will not enforce commercial contracts that stipulate ``manifestly unreasonable" delivery times.\footnote{See Uniform Commercial Code \S 28:1-204.} 

Second, negotiations are constrained by parties' outside options through ``default rules" in contract law, as well as a range of policies beyond contract law. For example, ``equitable division" rules for marriage contracts specify what will be enforced in the absence of a premarital agreement.\footnote{The ``equitable division" default stipulates that family courts split assets ``equitably" in the event of divorce in the absence of a premarital contract. The rule holds in 28 states. See, for example, Massachusetts General Laws Chapter 208 Section 34 for a statement of factors used in determining an ``equitable division" of assets.} Meanwhile, policies such as public healthcare influence workers' outside options in negotiations with their employers. 


Limits on enforceable contracts and default outcomes are used by regulators to achieve a variety of social objectives. As a salient contemporary example, consider the regulation of platform labor contracts in the United States, where regulators have articulated social objectives that include concerns about efficiency, equity and externalities. Secretary of Labor Marty Walsh spoke to the need to balance \textit{efficiency and equity} in early 2021: ``These companies are making profits and revenue and I’m not (going to) begrudge anyone for that... But we also want to make sure that success trickles down to the worker."\footnote{\url{https://www.washingtonpost.com/business/2021/04/29/labor-walsh-gig-workers-employees/}} Meanwhile, California's Assembly Bill 5 (AB5), which attempted to reclassify platform workers as employees in 2019, cited the \textit{externalities} that may accompany misclassification. The court noted the potential harm borne by taxpayers due to ``the loss to the state of needed revenue from companies that use misclassification to avoid obligations such as payment of... Social Security, unemployment, and disability insurance."\footnote{Assembly Bill 5 was signed into law in September 2019 only to be largely overridden by a statewide referendum, Proposition 22, in November 2020.}


This paper studies how governments optimally set immutable rules and default outcomes to promote efficiency, curb externalities, and achieve distributional objectives in contracting. We consider a setting in which the government is uninformed about a payoff-relevant state of the world that is common knowledge to two contracting parties. The parties have quasilinear preferences over ``quality" and monetary transfers, and bargain efficiently and costlessly over contract terms. The government’s goal is to maximize a \textit{generalized social welfare function} that weights efficiency, externalities and distributional concerns to some degree. The government chooses the social welfare-maximizing \textit{default delegation} mechanism, which defines a \textit{delegation set} of enforceable contracts, as well as a \textit{default outcome} which serves as a disagreement point in the parties’ negotiations.


The setting we introduce raises two distinct questions. First, under what conditions does default delegation achieve the social welfare-maximizing contract in every state? In order to answer this question, we take an implementation-theoretic approach to characterizing when default delegation implements general social welfare functions. The second question is more applied: How does the optimal default delegation policy respond to shifts in regulatory goals and empirical facts about the contract of interest? To study this question, we solve for the optimal default delegation policy in a setting where it is not without loss to restrict to default delegation.

The first insight from our analysis is that defaults and delegation sets serve distinct regulatory goals: defaults help the government achieve particular distributions of surplus while delegation sets internalize externalities. 
For example, we find that under a restrictive supermodularity condition on the agents' preferences, any social welfare function that concerns only efficiency and equity can be implemented by setting a default without restricting the delegation set. Similarly, restricting the delegation set implements social welfare functions that concern only efficiency and externalities. These characterizations, together, provide a limited mechanism design foundation for default delegation.



Our second contribution is an applied analysis of default delegation in the context of platform work. 
We perform comparative statics to understand how the optimal default delegation policy responds to changes in: (i) the government's social welfare function, (ii) worker bargaining power, and (iii) the composition of the platform workforce. As more workers treat platform work as a full-time job and as externalities from platform labor increase, optimal default delegation increases in-kind support for workers through improved outside options (e.g. social insurance) and limitations on enforced contracts (e.g. minimum required benefits). However, when worker bargaining power decreases away from parity, optimal default delegation improves worker's outside options through monetary transfers (e.g. unemployment benefits, minimum wage laws).

These results help to organize the debate about whether platform workers are misclassified as independent contractors in the United States. When California's AB5 was largely overturned, a new classification for platform workers emerged. This new classification effectively carved out a middle ground category between full-time employees and contractors. Our results show how the optimal ``middle ground" classification depends on what the policy is responding to. If the change in policy is principally a response to an increase in concerns about the distribution of surplus or externalities arising out of labor contracts, then the most important feature of AB5 is the minimum benefits requirement. Meanwhile, if the change in policy is a response to a decline in worker bargaining power, then the most important feature to maintain in the middle ground classification is unemployment insurance and minimum wages.\footnote{More broadly, in light of recent macro trends such as declines in the labor share of income and worker bargaining power \cite{AutorDavid2020TFOT, SummersLawrenceH2020TDWP}, a regulator with equity and externality concerns optimally adjusts the contracting environment through minimum wages, unemployment insurance, and other cash transfers that improve outside options.} 

 After discussing related literature, we present our model in section \ref{sec:model}. Then, section \ref{sec:first_best} contains an implementation-theoretic analysis of default delegation. Section \ref{sec:application} derives comparative statics on the optimal default delegation policy in the context of platform work. Section \ref{sec:conclusion} offers further discussion of the policy implications of our analysis, and concludes.

\paragraph{Related literature.} This paper contributes to recent applied theoretical analyses: of the trade-off between commitment and flexibility in specific policy settings \citep[for example]{halac2018fiscal, halac2020commitment, kartik2021delegation, ali2022sequential}; of platform regulation \citep[for example]{rochet2003platform, loertscher2020digital, gomes2020regulating, kangcontracting}; and of social objectives beyond efficiency \citep[for example]{saez2016generalized, millner2020nondogmatic, dworczak2021redistribution}. The model itself draws on literatures on contracting and implementation with renegotiation, and on delegated decision-making in organizations. 

Cast in the incomplete contracts framework of \citet{hartmoore1988}, our model is closest to \citet{aghiondewatripontrey1994}, which focuses on how an initial contract might include provisions that govern the ex-post renegotiation process.\footnote{A paper that makes a similar point to \citet{aghiondewatripontrey1994} is \citet{chung1991incomplete}. Other papers that study the role of renegotiation in incomplete contracts include \citep{hermalinkatz1991, greenlaffont1992, rubinsteinwolinsky1992}.} 
Our analysis extends \citeauthor{aghiondewatripontrey1994}'s and differs in interpretation and emphasis. Rather than supposing that the two parties to a contract write their ex-post renegotiation provisions (default outcomes, allocation of bargaining power) into the initial contract, we assume that there is a regulator or social planner who has the authority to set and enforce default outcomes (and we treat bargaining power as exogenous). This shift in interpretation reflects that we are studying not the provisions of bilateral contracts themselves, but rather how lawmakers regulate bilateral contracts in the interests of the public. Our framework allows us to study how optimal defaults change with different planner preferences over efficiency, equity and externalities, whereas \citeauthor{aghiondewatripontrey1994} focus on whether, with some optimally-specified default and allocation of bargaining power, efficient investment is possible. 

In providing a limited mechanism design foundation for default delegation, we contribute to the literature on Nash implementation  \citep{maskin1999, moorerepullo1988}. In this context, the possibility of renegotiation restricts the set of implementable outcomes---any outcome that is not on the agents' Pareto frontier will be renegotiated to realize a Pareto improvement. \citet{maskinmoore1999} fully characterize implementability when renegotiation cannot be prevented. Extending their analysis, we add a characterization of implementability for specific social welfare functions concerned with the distribution of surplus between contracting parties.

      The model can also be viewed through the lens of delegation, though, unlike most of the delegation literature, analyzes delegation to \emph{two} agents. A special case of our model (when the regulator cares only about efficiency and externalities) reduces to a canonical delegation problem as introduced by \citet{holmstrom1977, holmstrom1984} and generalized in \citet{alonsomatouschek}. While \citet{martimortsemenov2008} considers a two-agent delegation problem in a legislative context when agents are asymmetrically informed, we assume agents have symmetric information. Delegation differs from standard mechanism design problems in that it assumes neither the principal nor any third parties (as in \citet{tirole1986hierarchies} and \citet{laffontmartimort1998}) can collect or disburse transfers to or from the agents. 
 
    Our analysis contributes to the study of tradeoffs between decentralization and centralization in organizational economics and mechanism design, thoroughly surveyed in \citet{mookherjee2006review}. A common motivating example in the delegation literature is the regulation of a monopolist \`{a} la \citet{baronmyerson1982}. In the extant delegation literature, the results about the optimality of interval delegation are used to make sense of the widespread use of price caps in regulatory settings. Our model offers a new interpretation of regulation-as-delegation: when the government's social welfare function incorporates distributional concerns, optimal regulation takes into account the bargaining process between a firm and its stakeholders.

\label{sec:intro}

\section{Model} 
\label{sec:model}
We present our general model in terms of an employment contract between a worker and a firm. The firm and the worker, more generally, can be understood to be any two parties with symmetric information negotiating the terms of a bilateral contract. In line with the delegation perspective of our model, we sometimes refer to the firm and the worker collectively as the ``agents," while the regulator in this setting is the ``principal." The principal aims to maximize social welfare by influencing the contracting environment and choosing which contracts to enforce. 

The state of the world, which governs agent preferences (and thus social welfare), is observable but unverifiable. Thus, the principal is limited in that it cannot enforce contracts that are contingent on the state. In line with the delegation literature, we rule out the possibility that the principal can make monetary transfers to or from the agents, and further assume that the principal cannot mandate transfers between agents (i.e. cannot obstruct side-payments). 


\paragraph{Contracts.} Agents bargain over the terms of a contract $(q,c) \in \Q \times \C \subseteq \R^2$. The term $q$ in the contract represents a dimension over which agents have possibly state-dependent valuations. We refer to $q$ as ``quality," though it could represent many non-price features of a contract. For example, $q$ might represent a particular benefit in an employment contract, such as the degree of health insurance coverage provided by the firm to the worker. Meanwhile, $c$ captures the money that will be transferred from one agent to the other. In an employment contract, $c$ is the compensation (salary, wages) paid to the worker by the firm. 

\paragraph{Preferences.} The regulator and the agents' utilities depend on the negotiated contract and on the state of the world. The state of the world is $\theta \in \Theta \subset \mathbb{R}$. 

The agents, the firm ($f$) and the worker ($w$), have quasilinear state-dependent utility functions over the outcome $(q,c)$:
$$\textbf{Firm: } U_f(q,c; \theta) = u_f(q; \theta)-c \hspace{4mm} \textbf{Worker: } U_w(q,c; \theta) = u_w(q; \theta)+c.$$
The principal's goal is to maximize a generalized social welfare function,
   \begin{equation}\label{eq:swf_def}\SWF(q,c;\theta) = \SWF(U_f, U_w, U_r; \theta) =  \underbrace{U_f + U_w}_{\textbf{``efficiency"}} - \underbrace{\beta(U_f - U_w)^2}_{\textbf{``equity"}} + \underbrace{\gamma U_r(q; \theta)}_{\textbf{``externality"}}\end{equation}
   where $\beta\geq 0$ and $\gamma\geq 0$ scale the magnitude of the social cost associated with equity and externalities, respectively. Equity concerns are represented by a quadratic penalty. The equity term is maximized when the worker and firm attain the same utility. This ``equal split" equity objective is assumed to simplify exposition.\footnote{It is straightforward to verify that the results presented here hold for broad set of equity objectives, e.g. any particular desired distribution of surplus $x$ and $1-x$ for $x\in [0,1]$. A discussion of the form of the inequity penalty appears in \autoref{sec:inequity_penalty}.} Externalities are represented by $U_r(q; \theta)$. For example, in an employment contract, the government may end up paying for health care that is not covered by an employer provided health insurance program. 
   
    For the remainder of the paper, when we refer to an arbitrary social welfare function, $\SWF,$ we are referring to a function of the form in \eqref{eq:swf_def}. The generalized social welfare function nests more specific social welfare functions. For instance, when $\beta = 0$ and $\gamma=0$, the regulator's objective simplifies to maximizing efficiency. When $\beta=0$ and $\gamma>0,$ the regulator additionally attempts to internalize externalities that may arise out of the contract---the corresponding social welfare function will be referred to as a $\gamma\text{-}\SWF$. When $\beta>0$ and $\gamma=0,$ the regulator trades off efficiency losses with equity gains---the corresponding social welfare function will be referred to as a $\beta\text{-}\SWF$. When both parameters $\gamma$ and $\beta$ are greater than zero, the regulator has three competing objectives, and chooses a policy that optimally compromises among them.

\paragraph{Information.} The agents have complete and symmetric information about the state $\theta.$ However, the state is not verifiable by the principal. 
All other details of the environment are common knowledge among the principal and the agents, including the format and outcome of bargaining.

\paragraph{Bargaining.} Agents bargain over the terms of the contract $(q,c)$ given an outside option $d = (q_d, c_d)$ and the state of the world $\theta$. We assume the agents Nash bargain, as a convenient reduced-form expression for an efficient bargaining process which result in a particular distribution of the surplus from bargaining---the worker has bargaining weight $\delta$ while the firm has weight $(1-\delta).$ 
The default contract which obtains in the event of a disagreement is $d$. That is, the bargained contract is given by
\begin{equation}\label{eq:bargaining_def}
    h(d; \theta) \equiv \arg\max_{q\in \Q,c\in \mathbb R}  ( U_w(q,c;\theta) - U_w(d;\theta))^\delta( U_f(q,c;\theta) - U_f(d;\theta))^{1-\delta}.
\end{equation}
We write $\tilde{h}((q,c),d;\theta)$ to refer to the maximand in \eqref{eq:bargaining_def}. Here $\Q$ refers to the set of enforceable quality levels.

\paragraph{The regulator's problem: Default delegation.} The principal's goal is to maximize social welfare. The principal maximizes social welfare through \textit{default delegation}. 

Default delegation is an indirect mechanism in which the regulator uses two tools to affect the contracting environment: (1) it chooses the \emph{delegation set} $\Q$ of quality levels it is willing to enforce and (2) it sets the \emph{default outcome} outcome $d$ which serves as a disagreement point in the agents' negotiations. That is, the regulator maximizes
\begin{equation}
    \max_{\Q, d} \E_\theta[\SWF((q(\theta),c(\theta));\theta)]
\end{equation}
with
\begin{align*}
(q(\theta),c(\theta)) \equiv \argmax_{q \in \Q, c\in \mathbb R} \tilde{h}((q,c),d; \theta),
\end{align*}
for all $\theta \in \Theta$. In \autoref{sec:first_best}, we will discuss the limited conditions under which 
any SWF that is implementable with an arbitrary mechanism is implementable with default delegation. In \autoref{sec:application}, we conduct a second-best analysis: assuming that the regulator is limited to default delegation mechanisms.

\paragraph{Timing of default delegation.} The regulator chooses a delegation set and default outcome $\{\mathcal{Q},d\}.$ Then, with common knowledge of the true state \(\theta\), the agents bargain over the terms of the contract $(q,c),$ with $d$ serving as the disagreement outcome, and $\Q \times \C$ defining the feasible bargaining outcomes. Then the contracts are signed and utilities are realized. 

\paragraph{Discussion of default delegation.} It is worth considering why regulators are unable to mandate particular transfers between the contracting parties---that is, in the language of default delegation, why the regulator limits $\Q$ (the set of quality levels it will enforce), but does not limit the set of transfer levels it will enforce. 

First, note that enforcement is not costless in practice. Furthermore, policy choices influence enforcement costs. As noted by  \citet{GlaeserEdwardL2001ARfQ}, governments often regulate undesirable activities by limiting those activities rather than taxing them. They do so because it is less costly to identify violations of limits or mandates, than to identify violations of some payment requirement. In our setting, it may be excessively costly or impossible to enforce particular payments between contracting parties if the parties can transact in cash. 

Relatedly, if regulators were to limit the payments between parties, the set of enforceable contracts would take the form of a menu with specific $(q,c)$ pairs. So the enforcement costs may compound: not only would the government need to enforce particular payments $c$ between parties, but also they would need to enforce particular payments $c$\textit{ contingent on a particular} $q$, and vice versa. For instance, employment law does not prescribe schedules of acceptable combinations of health insurance coverage and compensation levels. Instead, employment law mandates a minimum level of health insurance coverage. Although we do see minimum wage laws in practice, it is important to note that in our model setting a minimum wage is isomorphic to setting a default transfer. \footnote{In particular, the minimum wage shows up in default delegation as a function of the default in workers' negotiations and the set of acceptable quality terms $\Q$. To see this, let $c_d$ be the default hourly wage for nonexempt employees. Suppose that quality $q$ is the level of employer-sponsored health insurance coverage (fraction of expenses covered). Then, $\qubar$ is the minimum coverage amount for employer-sponsored health insurance which, in the U.S., is 60\%. Note that $\qubar,$ together with $(q_d,c_d)$ implicitly defines a minimum wage: when $\qubar$ binds, the transfer level $c(q_d, c_d, \qubar)$ is the minimum wage or conversely, the minimum wage implicitly defines the default transfer.} 

\section{First-Best Analysis}
\label{sec:first_best}

In this section, we characterize the set of social welfare functions that are implementable with default delegation. The goal of this analysis is twofold: first, we aim to understand the mechanism design foundations and limitations of default delegation, and second, we aim to derive intuition about \textit{how} default delegation helps the principal balance among competing objectives to promote efficiency, ensure equity and mitigate externalities in bilateral contracting. 

We begin by showing that social welfare functions with an equity component can, under limited conditions, be implemented when the regulator has the power to set a \textit{default} which serves as the outside option in parties' bargaining. Social welfare functions concerned with externalities arising from a bilateral contract can, under even more limited conditions, be implemented when the regulator has the power to restrict the set of enforceable contracts. 

This section concerns the implementation of social welfare functions of the form in \eqref{eq:swf_def} \textit{with default delegation} mechanisms.\footnote{Appendix \autoref{sec:unconstrained} compares the set of social welfare functions implementable with default delegation to the set of social welfare functions implementable with any unconstrained mechanism.} We say that default delegation \textit{implements} a social welfare function if it achieves the first-best outcome in every state. 
\begin{definition}[First-Best Outcomes]
The \emph{first best outcomes} for a social welfare function $\SWF$ and a state space $\Theta$ are $\{(q_\theta,c_\theta)\}_{\theta \in \Theta}$ where 
$$(q_\theta,c_\theta) \in \argmax_{q,c} \SWF(q, c; \theta).$$
\end{definition} 

\begin{definition}[Implementation with Default Delegation]
A social welfare function $\SWF$ is \emph{implementable with default delegation (``DD-implementable")} under state space $\Theta$ if there exists a $(\Q, d)$ such that in all states $\theta,$ the outcome of bargaining given default $d$ and delegation set $\Q$ coincides with the first-best outcome. That is, 
$$(q_\theta, c_\theta)= \argmax_{q\in \mathcal Q, c } \tilde{h}((q,c), d;\theta)$$
for all $\theta \in \Theta,$ where $(q_\theta, c_\theta)$ is the first-best outcome in state $\theta$.
\end{definition}
For brevity, we sometimes say that a $\SWF$ is DD-implementable if it is implementable with default delegation. Proofs of all propositions are in Appendix \ref{sec:proofs}.




\subsection{DD-implementation with equity concerns}

First, note that social welfare functions concerned with efficiency alone are
trivially implementable with default delegation. With full freedom of contract, the parties reach
the first-best outcome. However, it may be that in order to achieve the first best level of quality $q_\theta$, the agents' bargaining results in an undesirable distribution of the surplus.\footnote{In our examples, the undesirability of a particular distribution could be derived from ex ante contracting efficiency (as in incomplete contracts, see Example 1 in \autoref{sec:examples}) or from fairness or other social concerns about distribution (as in marriage and employment contracts, see Examples 2 and 3 in \autoref{sec:examples}).} When the regulator has distributional concerns, i.e. $\beta>0$ in \eqref{eq:swf_def}, the first-best in state $\theta$ is a single pair $(q_\theta, c_\theta)$ where $q_\theta$ is the efficient quality level and $c_\theta$ evenly distributes the surplus between the two parties.\footnote{Recall that this particular first-best split comes from the specific social welfare function in \eqref{eq:swf_def}, which is maximized when both parties have equal surplus. As noted in \ref{sec:model} and further explored in Appendix \ref{sec:inequity_penalty}, this choice of $\SWF$ simplifies exposition---the results presented here can be derived for any particular distribution of the surplus $x, (1-x)$ for $x\in[0,1]$.}

We first discuss DD-implementation of social welfare functions with \textit{only} efficiency and equity concerns---i.e. $\beta\text{-}\SWF$s. Our analysis focuses on the case in which $\Theta$ contains two states. In this case, it is without loss to restrict to default delegation mechanisms: any $\betaswf$ that can be implemented with an arbitrary mechanism with bargaining can be implemented with default delegation. In order to compare DD-implementability to general implementability, we draw on the following result from \citet{maskinmoore1999}.

\begin{proposition}[\citet{maskinmoore1999}]\label{prop:maskinmoore} Assume that the Pareto frontier is linear in all states $\theta \in \Theta$. A $\SWF$ is implementable with an arbitrary mechanism with bargaining in Nash equilibrium and any refinement if and only if there exists a function $g: \Theta \times \Theta \to \mathcal{Q} \times \mathcal{C}$ such that 
\begin{itemize}
\item[(i)] $$h(g(\theta, \theta), \theta) \in \argmax_{g(\theta, \theta)} \SWF(h(g(\theta,\theta), \theta)),$$
    \item[(ii)] and for $i \in \{w,f\}$ 
    \begin{equation}
   U_i(h(g(\theta, \theta), \theta)) \geq   U_i(h(g(\theta', \theta), \theta))
\end{equation}
for all $\theta, \theta' \in \Theta.$
\end{itemize}
\end{proposition}
Since agents' utility functions $U_f$ and $U_w$ are quasilinear, Maskin and Moore's result characterizes, at a high level of generality, which $\SWF$s are implementable with an arbitrary mechanism with bargaining in our model.\footnote{Note that Maskin and Moore's result holds for more general bargaining functions $h$ which satisfy assumptions of predictability, efficiency and individual rationality. As Nash bargaining satisfies these assumptions, the result applies directly in our context.}

To make Proposition \ref{prop:maskinmoore} more concrete, consider the two state case where \(\Theta = \{\theta_l,\theta_h\}\). Here, we write the first best outcomes to be $(q_l,c_l)$ and $(q_h, c_h)$ in state $\theta_l$ and $\theta_h,$ respectively.  Proposition \ref{prop:maskinmoore} says that the first best is implementable in any refinement of Nash equilibrium as long as neither player would deviate in the reduced form game shown in \autoref{tab:general_game_2states}.

   \begin{table}[H]
     \centering
        \begin{tabular}{c|c|c}
            & \(\thetahat_f=\theta_l\) & \(\thetahat_f=\theta_h\)   \\
            \hline 
            & & \vspace{-0.35cm} \\   
                \(\thetahat_w=\theta_l\) & \((q_l, c_l)\) & \((q_{lh},c_{lh})\) \\
                 \hline 
                 & & \vspace{-0.4cm} \\   
                 \noalign{\vskip 1mm}    
                \(\thetahat_w=\theta_h\) & \((q_{hl},c_{hl})\) & \((q_h,c_h)\) \\
                 \hline 
            \end{tabular}
     \caption{Direct mechanism for implementing $(q_\theta, c_\theta)$ in state $\theta$.}
     \label{tab:general_game_2states}
 \end{table}
The regulator's problem then is to choose a mechanism $g(\thetahat)$ so that the only equilibria of the game are $(q_\theta, c_\theta)$ for $\theta \in \Theta.$ This amounts to a selection of off-path ``threats" 
$(q_{hl},c_{hl})$ and $(q_{lh},c_{lh})$. In selecting the optimal threats the regulator anticipates, the negotiation that would take place in the event that the off-path outcome were to take place. For instance, consider what would happen if \(w\) deviates in state \(\theta_l\) by choosing \(\hat{\theta}=\theta_h\). The worker and firm would then negotiate from $(q_{hl},c_{hl})$ to $(q_{l},c_{l})$. This would result in the worker receiving utility: 
\[u_w(q_{hl}; \theta_l)+\delta \left(\sum_i u_i(q_l,\theta_l) - \sum_i u_i(q_{hl},\theta_l)\right)+c_{hl}\]    

In Appendix \ref{sec:proofs}, we show that in order to prevent the worker and the firm from deviating to the outcome $(q_{hl},c_{hl})$, it must be that 

\begin{equation}\label{eq:ic_lh}
(1-\delta)[\Delta\Delta(u_w, q_{hl}, \theta_h, \theta_l)] - \delta[\Delta\Delta(u_f, q_{hl}, \theta_h, \theta_l)]\geq c_l - c_h
\end{equation}
where we define a double-difference operator: $\Delta\Delta(u, q,\theta, \theta') \equiv [u(q_\theta, \theta) - u(q, \theta)] - [u(q_{\theta'}, \theta') - u(q, \theta')].$ An analogous condition must hold in order to prevent deviations to $(q_{lh},c_{lh})$. In order to gain some intuition for the double-difference operator, note that each term in brackets represents the utility gained by moving from some arbitrary \(q\) to the first best in a particular state. Thus, the double-difference operator is the difference between these gains across two states $\theta$ and $\theta'.$ By analyzing the reduced form game in \autoref{tab:general_game_2states}, we prove the following proposition.

\begin{proposition} \label{prop:imp_default} Assume that $|\Theta|=2$. If a social welfare function $\betaswf$ is implementable with an arbitrary mechanism with bargaining, and $U_f$ and $U_w$ are continuous, then the $\betaswf$ is DD-implementable. Letting $\Theta=\{\theta_\ell, \theta_h\}$, the default quality level $q_d$ that implements the $\betaswf$ satisfies 
\begin{equation}\label{eq:default_cond}
    (1-\delta)[\Delta\Delta(u_w, q_d, \theta_h, \theta_\ell)] - \delta[\Delta\Delta(u_f, q_d, \theta_h, \theta_\ell)] = c_\ell - c_h.
\end{equation}
\end{proposition}

To gain intuition for condition \eqref{eq:default_cond} in Proposition \ref{prop:imp_default}, consider the case where \(\delta=0\). In this case, condition \eqref{eq:default_cond} states that if the workers gain more from bargaining over quality in the high state than in the low state (i.e. the left hand side of condition \eqref{eq:default_cond} is large), then they are willing to accept a lower transfer in the high state. When \(\delta \neq 0\), the wedge created by the workers bargaining from the default is offset by the differential gains that the firm accrues. Condition \eqref{eq:default_cond} emphasizes that DD-implementability of $\betaswf$s depends on the existence of a default value $q_d$ that is valued differently across states by the agents. To give further insight into the role of state-dependence, the following corollary presents a sufficient condition for DD-implementability of $\betaswf$s. 

\begin{corollary}\label{cor:const_super} Assume $|\Theta|=2$. Then if there exists $x \in \mathbb{R}_{+}$ such that at all points $(q, \theta)$ 
\begin{equation}\label{eq:super_mod}
     \left|(1-\delta)\frac{\partial^2 U_w}{\partial q \partial \theta} - \delta \frac{\partial^2 U_f}{\partial q \partial \theta}\right| > x,
\end{equation}
any $\betaswf$ is DD-implementable with a default $d$.
\end{corollary}
This sufficient condition shows that if agents' utility functions are sufficiently---and differentially---responsive to the state any $\betaswf$ is DD-implementable. Note a special case: if the worker has supermodular preferences and the firm has submodular preferences then any $\betaswf$ is DD-implementable (because the LHS of \eqref{eq:super_mod} is always greater than 0). 

In the context of our main example, if workers value additional benefits more in the high state of the world than in the low state of the world, low default benefits means the workers will benefit more from bargaining in the high state than in the low state. Thus, the transfer they receive in the high state after bargaining will be relatively low. The opposite is true for a sufficiently high default level of benefits. Corollary \ref{cor:const_super} states that in this situation if \(\delta =0\), any $\betaswf$ is DD-implementable. Note that this implies that a large difference in transfers \(|c_\ell - c_h|\) would require a sufficiently extreme default. 

However, there may be instances in which such an extreme default is not feasible. An interesting implication of Corollary \ref{cor:const_super} is that the bargaining parameter can have a significant impact on whether a $\betaswf$ is DD-implementable and what value of \(q_d\) implements a particular $\betaswf$. As the bargaining power of the party with \textit{more state-dependent preferences} decreases, the \(q_d\) which would achieve a particular difference in transfer becomes less extreme. This result is stated with additional assumptions for the purposes of clarity in Corollary \ref{cor: bargaining_and_default}

\begin{corollary}\label{cor: bargaining_and_default}
Assume $|\Theta|=2$. Fix $U_f, U_w,$ and $\Theta$ and suppose the worker and firm have preferences such that 
\[\frac{\partial^2 U_w}{\partial q \partial \theta} = b > 0 \hspace{3mm}\text{  and   }\hspace{3mm} \frac{\partial^2 U_f}{\partial q \partial \theta} = -a < 0\]
and \(b>a\). Suppose default $\hat{d}$ DD-implements a particular $\betaswf$ at $\hat{\delta}.$ Then for all $\delta < \hat{\delta}$, if \(d\) DD-implements the same $\betaswf$ at \(\delta\), then \(|q_d-q_{\theta_h}|<|\hat{q}_d-q_{\theta_h}|\).
\end{corollary}
Corollary \ref{cor: bargaining_and_default} is somewhat counterintuitive. Loosely put, it states that when the workers' preferences over \(q\) are more responsive to the state $\theta$, the regulator can more ``easily" implement a $\betaswf$ when the firm has higher bargaining power. The intuition behind this result will reappear in the analysis in \autoref{sec:application}.


\subsection{DD-implementation with externality concerns}


Now, we turn to social welfare functions with externalities, i.e. SWFs with $\gamma>0$.  In such cases, the first-best $(q_\theta, c_\theta)$ is not on the agents' Pareto frontier, and so the regulator will not be able to attain the first best without restricting the set of enforced contracts $\Q$. To simplify the analysis, we first consider DD-implementability of $\gammaswf$s (in which $\gamma>0$ but $\beta=0$).

If the externality depends on the quality level such that \(\frac{\partial U_r}{\partial q}\neq 0\) and $\gamma>0$, the regulator may still be able to DD-implement the $\gammaswf$ by delegating the choice to the agents; i.e. letting them choose from a reduced set of contracts that contain the optimal quality levels corresponding to the first-best outcomes in each state. In other words, a $\gammaswf$ may be implementable with delegation set $\Q = \{q_\theta\}_{\theta \in \Theta}$. The following proposition characterizes DD-implementable $\gammaswf$s.
\begin{proposition}\label{prop:fb_with_gamma}
A 
$\gammaswf$ is DD-implementable under type space $\Theta$ if \begin{equation}\label{eq:fb_with_gamma}
        u_f(q_\theta, \theta) + u_w(q_\theta, \theta) \geq  u_f(q_{\theta'}, \theta) + u_w(q_{\theta'}, \theta) 
    \end{equation}
    for all $\theta, \theta' \in \Theta,$ and for all $q_\theta, q_{\theta'}\in \Q.$
    The delegation set is $\Q = \{q_\theta\}_{\theta \in \Theta}$.
\end{proposition}
Condition \eqref{eq:fb_with_gamma} is restrictive---in each state the agents' total surplus must be higher at the first-best level of quality, $q_\theta$, than at any other quality level $q_{\theta'}.$\footnote{Recall that we have assumed that when the Nash bargaining solution is not in the feasible set, the agents select the point in the feasible set that maximizes their joint surplus.} 

We can combine the previous results for the two state case (Propositions \ref{prop:imp_default} and \ref{prop:fb_with_gamma}) to get a more general statement about DD-implementation. We get the general result: 

\begin{proposition}\label{prop:fb_with_gamma_beta} Assume $|\Theta|=2.$ Then a social welfare function $\SWF$ is DD-implementable if 
\begin{itemize}
    \item[(i)] there exists $d=(q_d, c_d)$ that satisfies \eqref{eq:default_cond}, and 
    \item[(ii)] $\Q = \{q_\theta\}_{\theta \in \Theta}$ satisfies \eqref{eq:fb_with_gamma} for all $\theta, \theta' \in \Theta.$
\end{itemize}
\end{proposition}
 Now that we have discussed the conditions under which default delegation implements $\betaswf$s and $\gammaswf$s, we next turn to a discussion of the limitations of default delegation. 

\subsection{Limitations of default delegation}


When there is bargaining, it is without loss to restrict to default delegation when there are only two states and either no externalities or no equity concerns. Proposition \ref{prop:imp_default} says that any $\betaswf$ implementable with an arbitrary mechanism with bargaining is implementable with default delegation under mild conditions. Meanwhile, Proposition \ref{prop:fb_with_gamma} shows general conditions for implementing a $\gammaswf$. To the extent that real-world settings can be approximated by a binary state space, Propositions \ref{prop:imp_default} and \ref{prop:fb_with_gamma} provides implementation theoretic foundation for the use of default delegation. \footnote{Some settings which appear to have non-binary state spaces may nonetheless reduce to a binary state case. For instance, when the regulator has a maxmin objective function, the analysis in Proposition \ref{prop:imp_default} applies. See  Appendix \ref{sec:maxmin} for discussion.} 

However, when there are more than two states or the regulator has both equity and externality concerns, default delegation will, in general, meaningfully constrain the regulator's ability to achieve first-best relative to an arbitrary mechanism.\footnote{See Appendix \ref{sec:threestates} for further discussion.} Nonetheless, default delegation may be valuable in a second-best world, due to costs associated with the implementation of a more general mechanism. For instance, more general mechanisms may require a high degree of centralized communication, and complex enforcement schemes. Furthermore, by definition, default delegation achieves higher expected welfare than two benchmark alternatives: (i) total freedom of contract and (ii) mandating the ex-ante optimal contract. 

When we turn to a more in-depth second-best analysis of default delegation in \autoref{sec:application}, the logic of the binary state case will serve as an insightful guide: defaults help the regulator achieve its equity goals, while limits on the delegation set help the regulator achieve its goal of mitigating externalities. The limitations of default delegation will also become clear as the optimal policy will be complicated by interactions between the optimal default and delegation set. 

\section{Second-Best Analysis: An Application to the Regulation of Platform Work}
\label{sec:application}
In this section, we characterize the expected welfare-maximizing default delegation policy assuming particular functional forms for preferences. This approach will allow us to derive comparative statics providing additional intuition for the use of default delegation. We discuss the implications of these results in more depth in the context of regulating platform work in \autoref{sec:conclusion}. 


\subsection{Set up}

Consider a regulator concerned about the contracts written between a continuum of firms and their app-based workers. The regulator is uncertain about how any given worker values benefits $q$ relative to cash $c$. To fix ideas, let $q$ be the level of health insurance coverage. 
We capture the regulator's uncertainty about worker's valuation for coverage through $\theta$, an unknown state of the world drawn independently from a distribution \(G(\theta)\) with support $ [\underline{\theta},\overline{\theta}]$. 

Although $\theta$ is unknown to the regulator, it is common knowledge to each worker-firm pair. Each worker's preference is quadratic and reaches a maximum at $q=\theta$. Some workers put relatively high value on additional  health insurance coverage relative to additional pay while others receive relatively little additional surplus from additional benefits.\footnote{See \cite{NBERw29736} for survey data reporting substantial heterogeneity in gig-worker's valuations of benefits.}

The firm's costs are state-independent: $q^2$ is the known cost of providing benefits. These costs are convex, and the firm always prefers to provide the minimum level of coverage. To summarize, workers and the firm have preferences
\begin{equation*}
\textbf{Firm: } U_f(q,c) = R - q^2 -c \hspace{5mm} \textbf{Worker: } U_w(q,c; \theta) = c -(q-\theta)^2  \\ 
\end{equation*}
where \(R\) is the revenue generated by the worker-firm pair.

We model the externalities generated by the contract as linear in $q$, for ease of exposition. (The results presented in this section would be qualitatively similar for increasing functions of $q$.) Since the externality term is linear, the socially optimal level of $q$ is always above the agent-optimal quality level. This externality term captures the fact that the regulator would be forced to cover health care that is not covered in the employer provided health insurance. 
\begin{equation*}
\textbf{Externality: } U_r(q) = q 
\end{equation*}
 All of these terms enter into an $\SWF, $ that is a generalized social welfare function of the form in \eqref{eq:swf_def}, with equity parameter $\beta$ and externality parameter $\gamma.$

\paragraph{The regulator's program.} In a second-best environment, the regulator aims to maximize \emph{expected} welfare. The regulator maximizes expected social welfare by choosing a default delegation policy to solve the following program: 

\[\max_{\{d, \mathcal{Q}\}} \E_\theta[\SWF(q_\theta, c_\theta;\theta)]\quad \text{such that} \quad q_d \in \mathcal{Q}\]
and
\[(q_\theta,c_\theta) = \argmax_{{q\in \mathcal{Q}}, c\in \mathbb{R}} (U_w(q,c;\theta) - U_w({q_d, c_d};\theta))^\delta(U_f(q,c) - U_f({q_d, c_d)})^{1-\delta}.\]
That is, the regulator chooses a delegation set $\mathcal Q$ and default $d=(q_d, c_d)$ to maximize expected welfare with respect to its prior $G(\theta)$, foreseeing that the agents will Nash bargain. The default quality level must lie in the delegation set (\(q_d \in \mathcal Q\)). 



In order to solve the regulator's program, we employ results from single-agent delegation problems. Note that conditional on a particular default, the regulator interacts with the agents as if they were a single combined agent (maximizing their joint surplus). \citet{alonsomatouschek} characterizes settings in which the optimal delegation set for a single agent is an \textit{interval}. Building on their results, we restrict attention to cases in which $\mathcal{Q}$ is an interval, i.e. $\mathcal{Q}= [\underline{q}, \bar{q}]$.\footnote{We discuss the use of interval delegation in more detail in Appendix \ref{sec:AM_application} and show conditions under which the optimal default delegation mechanism will take the form of a closed set. }


Let $(q_\theta, c_\theta)$ be the contract that the agents choose conditional on $(\mathcal Q, d)$ when the state of the world is $\theta.$ (This is a slight shift in notation from \autoref{sec:first_best}, where $(q_\theta, c_\theta)$ represented the first-best contract in state $\theta$.) Note that $\frac{\theta}{2}$ is the value of $q$ that maximizes the joint surplus of the worker and the firm. The agents bargain to the value in the delegation interval that maximizes their joint surplus. If the joint surplus-maximizing value of $q$ is not in the delegation interval, then $q_\theta$ is an endpoint of the interval. 

Using the fact that the optimal delegation set is an interval, and assuming that the worker and firm bargain to the value $q_\theta$ in the delegation set that maximizes their joint surplus, we can rewrite the regulator program: 
\begin{equation*}
    \max_{\{d,[\underline{q},\overline{q}]\}} \int_{\underline{\theta}}^{\overline{\theta}} \left(U_w(c_\theta, q_\theta; \theta) + U_f(c_\theta, q_\theta)+\gamma q_\theta - \beta\left(U_w(c_\theta, q_\theta; \theta) - U_f(c_\theta, q_\theta)\right)^2\right)dG(\theta)
\end{equation*}
\begin{align*}
   & \text{subject to} \\
  (q_\theta, c_\theta) & = \begin{cases} 
  (\underline{q}, c(\underline{q},d;\theta)) & \theta < 2\underline{q} \\ 
  (\frac{\theta}{2}, c(\frac{\theta}{2},d;\theta)) & 2\underline{q} \leq \theta \leq 2\overline{q} \\
   (\overline{q}, c(\overline{q},d;\theta)) & \theta > 2\overline{q}
  \end{cases}
\end{align*}
where $c_\theta$ is an implicitly defined function
\begin{equation}
c_\theta = c(q_\theta,d;\theta) = c_d +(1-\delta)(-(q_d -\theta)^2+(q_\theta-\theta)^2))-\delta(-q_d^2+q_\theta^2)).
\end{equation}


In the remainder of this section, we solve the regulator's program in three different bargaining regimes. As the regulator's program is sensitive to the exogenous bargaining parameter $\delta$, we simplify the analysis by considering three cases: equal bargaining (\(\delta = .5\)), firm control (\(\delta = 0\)) and worker control (\(\delta = 1\)). In order to get explicit solutions, we will sometimes make specific assumptions about the distribution of types (e.g. that the regulator's prior is uniformly distributed on the unit interval, i.e. $G(\theta) = \operatorname{Unif}[0,1]$).

\subsection{Firm control}

We begin our second-best analysis with the case in which the workers have no bargaining power, i.e. $\delta=0$. In this case, the firm makes take-it-or-leave-it offers to the workers. This assumption on $\delta$ aligns best with the facts about gig work in the U.S. Since gig-workers are not classified as employees, they are not protected by collective bargaining laws and so cannot form unions. Furthermore, factors that are not specific to gig-work contribute to low worker bargaining power in the U.S.: declining unionization \citep{farberetal2021}, rising employer concentration \citep{benmelech2020strong}, and diminished worker protections all contribute to low levels of worker bargaining power across sectors \citep{SummersLawrenceH2020TDWP}.  

\begin{figure}[h!]
    \centering
 \includegraphics[scale=0.45]{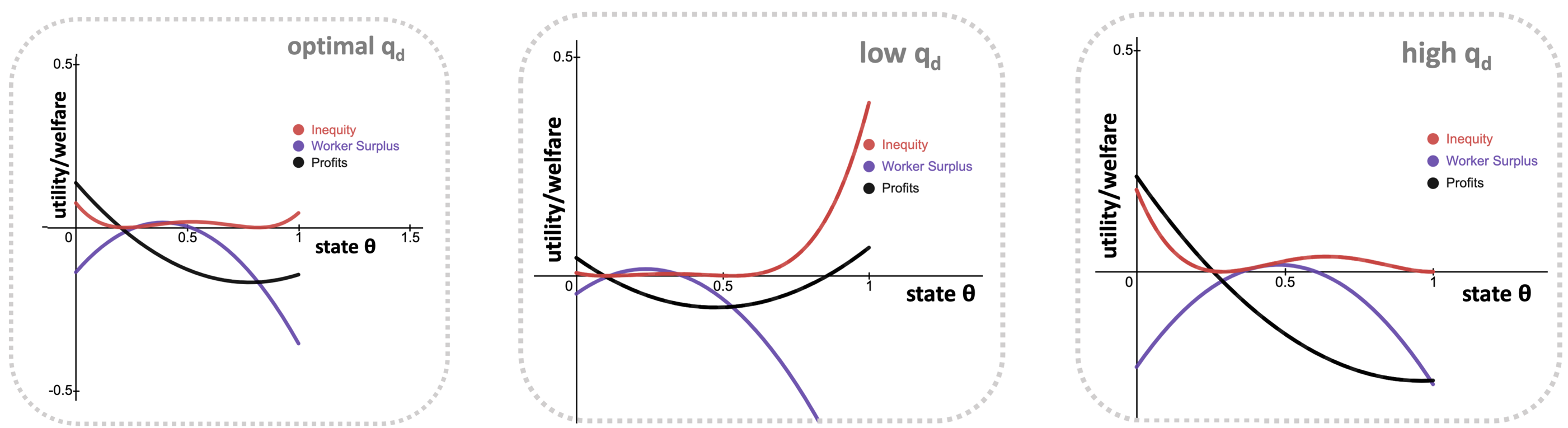}
\footnotesize
    \caption{\footnotesize Distribution of Surplus with Different Defaults. Each graph plots the worker's utility, firm's profits, and the inequity term in the social welfare function at different values of $q_d$, as a function of the state, $\theta$. When \(q_d = q_d^*\), the level of inequity is minimized across states (left). A lower default quality level $q_d < q_d^*$ exacerbates inequity in the high states while mitigating it in the low states (middle). A higher default quality level $q_d > q_d^*$ exacerbates inequity in low states while mitigating it in the high states (right). }
    \label{fig:opt}
\end{figure}

\paragraph{Solving the regulator problem.} When $\delta=0,$ the welfare-maximizing default delegation policy is complex. In particular, the endpoints of the delegation set \(\underline{q}\) and \(\overline{q}\) as well as the default \(q_d,c_d\) are interdependent. Each is a function of the other terms, as well as the equity and externality parameters \(\beta\) and \(\gamma\). To get a closed-form solution, we assume that types are distributed uniformly on the unit interval.
\begin{assumption}\label{ass:uniform_unit}
Types $\theta$ are drawn from the uniform distribution, \(\theta \sim \operatorname{Unif}[0,1]\). 
\end{assumption}
We first consider a case in which the maximum \(\overline{q}\) and minimum \(\underline{q}\) do not constrain bargaining in any state of the world.\footnote{This occurs when $\beta$ and $\gamma$ are both small.} 
\begin{assumption}\label{ass:unconstrained}
The outcome of bargaining $q_\theta$ is $\frac{\theta}{2}$ in all states. 
\end{assumption}
Under Assumptions \ref{ass:uniform_unit} and \ref{ass:unconstrained}, we get the following expected welfare-maximizing default: 

\begin{equation}
   \label{eq:delta0_default_unconstrained}
   q_d^* = \frac{3}{8},\quad c_d^* = \frac{R}{2}+\frac{1}{64}.
\end{equation}
The default transfer $c_d^*$ is above the level which equates the exogenous components of surplus, \(\frac{R}{2}\). Similarly, the default quality level $q_d^*$ is closer to the worker's expected optimal quality level, \(q=\frac12\), than the firm's optimal, \(q = 0\). This default illustrates the following, more general observation.
\begin{result}\label{res:delta0_d}
When the firm has all of the bargaining power $(\delta=0)$, the welfare-maximizing default $ (q_d^*, c_d^*)$ favors the worker in terms of both the quality level and the transfer. 
\end{result}
In order to equalize the surplus between the two parties, the equity-concerned regulator uses the default to improve the worker's position before bargaining occurs. See \autoref{fig:opt}.


\subsubsection{Minimums mitigate externalities and influence the default}


Recall from the results in \autoref{sec:first_best} that the main reason the regulator constrains the delegation interval is to internalize externalities. In this example, since there is a positive externality associated with increased health insurance, the regulator imposes a minimum quality level when the value of the externality is large. This, in turn, influences the optimal default. When Assumption \ref{ass:unconstrained} does not hold, the optimal default quality is a function of the minimum and maximum: 
\begin{equation}
q_d^*=\frac{1+2\overline{q}-8(1-\overline{q})\overline{q}^3+8(1-\underline{q})\underline{q}^3}{4}.    
\end{equation}
The impact of constraining the delegation set on the optimal default quality level is summarized in the following result.

\begin{result}\label{res:qd}
The default quality level $q_d^*$ is increasing in the minimum quality level $\qubar$ and decreasing in the maximum quality level $\qbar$. 
\end{result}

Result \ref{res:qd} might seem counter-intuitive because a policy which favors workers should be balanced by a policy which favors the firms if equity concerns are important. However, an increase in the minimum benefits reduces the extent to which an increase in the default leads to higher inequity by preventing unequal bargaining from the high default to a value below the minimum. Thus, the default quality level can be raised in order to reduce inequity in the high states as discussed with regard to the right graph in Figure \ref{fig:opt}. The equity concerns in this case are balanced through a decrease in the optimal transfer as will be shown more clearly below. 

\subsubsection{How changes in the social welfare function affect default delegation}

Results \ref{res:delta0_d} and \ref{res:qd} give insight into how the optimal default $(q_d^*, c_d^*)$ is set when the firm has all the bargaining power. We next look at how $(q_d^*, c_d^*)$ changes with the parameters in the regulator's social welfare function. 

Governments' preferences over equity and externalities are not fixed. The social cost of inequities between different parties, captured by $\beta$, changes over time as social preferences shift and as different political parties transition into and out of power. Externalities from a particular type of contractual arrangement also change as the composition of people and firms entering that arrangement shift over time. 

\begin{figure}[h]
    \centering
    \begin{minipage}{0.3\textwidth}
        \centering
\includegraphics[scale=.25]{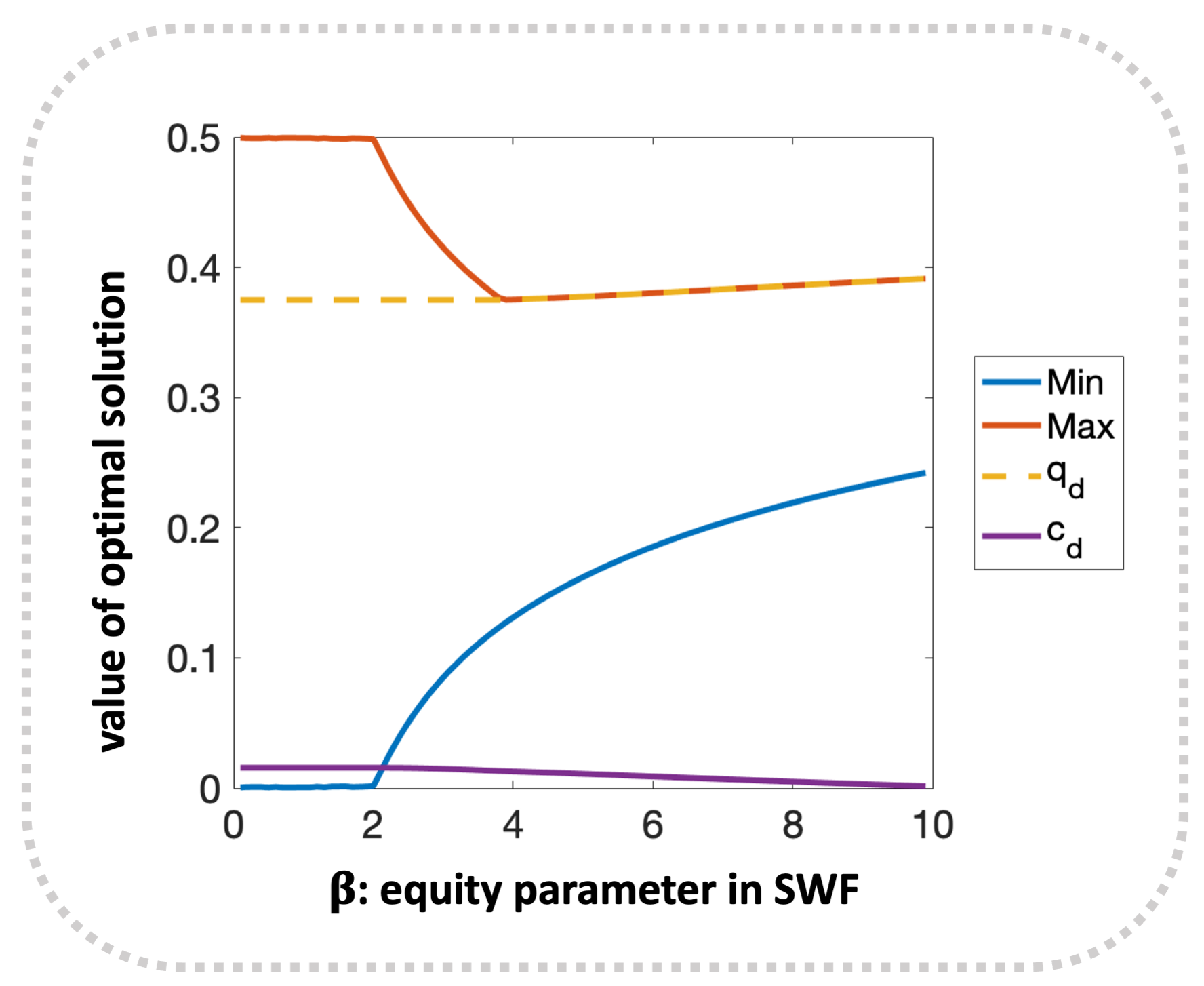}
\subcaption{$\delta=0, \gamma=0$}
    \end{minipage}\hfill
        \begin{minipage}{0.3\textwidth}
        \centering
\includegraphics[scale=.25]{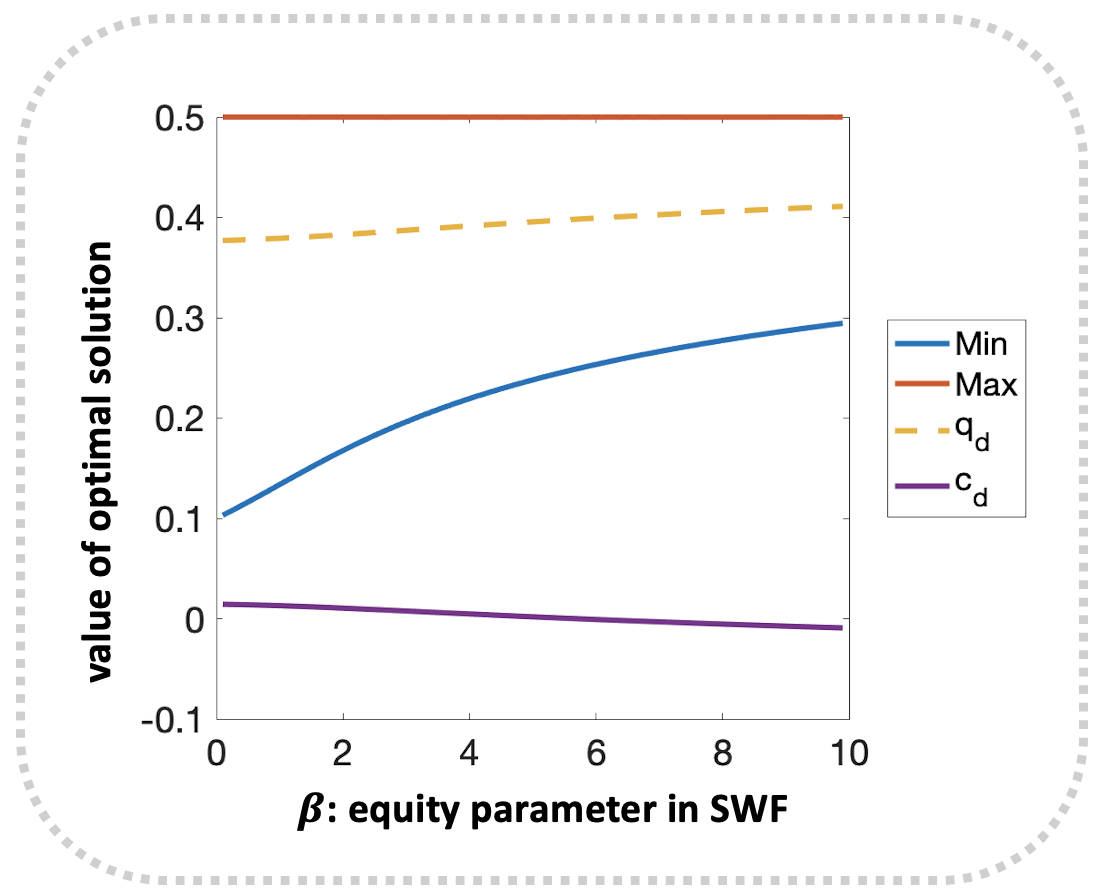}
\subcaption{\footnotesize $\delta=0, \gamma=.2$}
    \end{minipage}\hfill
    \begin{minipage}{0.3\textwidth}
        \centering
\includegraphics[scale=.25]{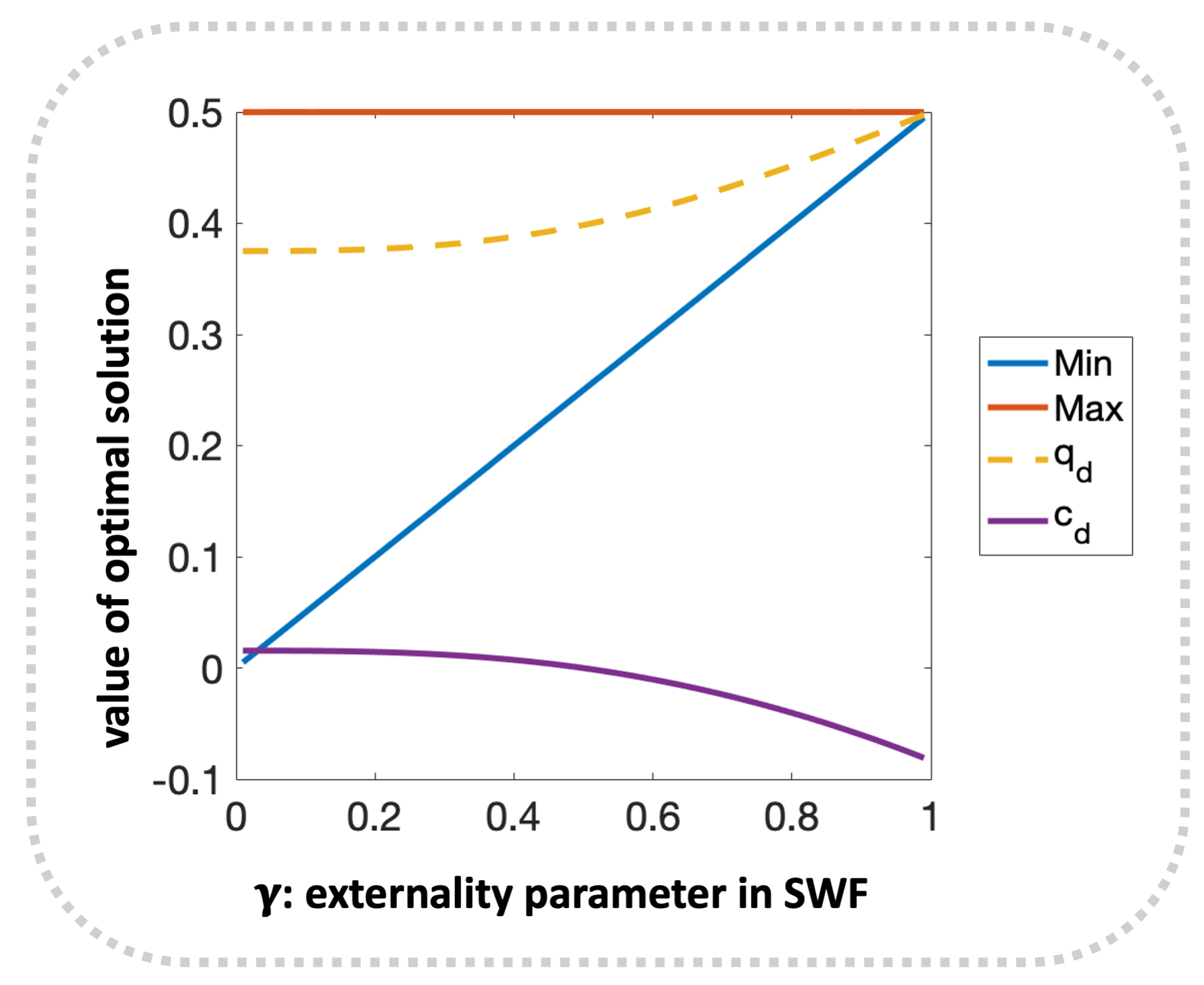}
\subcaption{\footnotesize $\delta=0, \beta=0$}
    \end{minipage}
    \caption{\footnotesize How Optimal Default Delegation Changes as Equity $(\beta,$ left, middle) and Externality ($\gamma$, right) Concerns Increase. Assume Assumption \ref{ass:uniform_unit} holds. \textbf{Panel (a)} plots the optimal default delegation policy $(\Q^*, d^*)$ as a function of the inequity penalty parameter ($\beta$), assuming the $\SWF$ is a $\betaswf$. 
    At high levels of $\beta,$ the maximum, minimum and default converge such that they effectively mandate the ex-ante most equitable contract. 
    \textbf{Panel (b)} plots the optimal default delegation policy $(\Q^*, d^*)$ as a function of the inequity penalty parameter ($\beta$), assuming the $\SWF$ has $\gamma=.2$; \textbf{Panel (c)} plots $(Q^*, d^*)$ as a function of $\gamma$ in a $\gammaswf$. 
    }

        \label{fig:2}

\end{figure}

We first consider how the optimal default delegation policy changes with the size of the inequity penalty $\beta$. As soon as $\beta$ is strictly positive, the first-best features an equal distribution of surplus. Panel (a) in \autoref{fig:2} plots the optimal default delegation policy for the $\betaswf$ in our example, and shows that for $\beta \in (0,2)$ the optimal default $q_d^*$ is constant at the value in \eqref{eq:delta0_default_unconstrained}. We call the value of $\beta$ beyond which equity concerns dominate efficiency concerns the \emph{equity-efficiency threshold}. In this example, the equity-efficiency threshold is $\beta=2.$  

\begin{result}\label{res:beta_compstat}
When the firm has full bargaining power ($\delta=0$), as equity ($\beta$) concerns increase, 
\begin{itemize}
    \item[(i)] up to the equity-efficiency threshold, the optimal delegation interval $[\qubar^*, \qbar^*]$ and default quality $q_d^*$ are constant, and
    \item[(ii)] beyond the equity-efficiency threshold, $[\qubar^*, \qbar^*]$ shrinks to reduce the unequal gains from bargaining.\footnote{This is one of main differences between the first-best and second-best analysis: in the first-best analysis, Proposition \ref{prop:imp_default} shows that the regulator cannot do better from an equity perspective by constraining the set of enforced contracts. However, as most cases of interest will feature levels of $\beta$ that are below the equity-efficiency threshold, the intuition that the regulator constrains the interval mainly to curb externalities holds.} The default quality $q_d^*$ responds following Result \ref{res:qd}, and the default transfer $c_d^*$ compensates this adjustment. 
\end{itemize}
\end{result}
This result highlights an important distinction between the default quality level $q_d^*$ and the default transfer $c_d^*$. Even as equity concerns increase past the equity-efficiency threshold, the default transfer $c_d^*$ is decreasing. Thus, the regulator is \textit{not} simply trying to improve the worker's position vis-a-vis the firm. Rather, the regulator is balancing the value that the workers receive in the form of health insurance against the transfers that they receive. Since workers have state-dependent preferences with respect to health care, the regulator can reduce the inequity across states by transferring more surplus to the workers in the form of default health care $q_d^*$, while reducing the transfers they receive. 


Next we consider how the optimal default delegation policy changes with the size of the externalities. Panel (c) in \autoref{fig:2} plots the optimal default delegation policy for a $\gammaswf$ as a function of $\gamma$. As the externality term \(\gamma\) increases, the minimum \(\underline{q}\) increases. This straightforwardly highlights the primary rationale for limiting the interval of enforced contracts as introduced in Proposition \ref{prop:fb_with_gamma}: restricting the interval of enforced contracts mitigates externalities. 

\begin{result}\label{res:gamma_compstat}
When the firm has full bargaining power ($\delta =0$), as externality concerns ($\gamma$) increase, the optimal default-delegation policy restricts the delegation interval $[\qubar^*, \qbar^*]$ in the direction which internalizes the externality. The default quality $q_d^*$ responds to the change in $\qubar^*$ or $\qbar^*$ following Result \ref{res:qd}, and the default transfer $c_d^*$ compensates this adjustment.
\end{result}

While the results for $\betaswf$s and $\gammaswf$s are for the most part intuitive, the strength of this model lies in its power to systematically analyze how default delegation achieves externality and equity objectives simultaneously. When \(\gamma>0\), any increase in equity concerns will lead to an increase in the optimal minimum as shown in the middle panel of Figure \ref{fig:2}. When there are externalities, the minimum is already balancing efficiency costs with the benefits of internalizing the externality. Thus, an increase in equity concerns simply tilts the scale further in favor of raising the minimum.

\begin{result}\label{res:beta_gamma_compstat} 
When the firm has full bargaining power ($\delta =0$), as equity concerns ($\beta$) increase in the presence of externality concerns, the optimal default-delegation policy restricts the delegation interval $[\qubar^*, \qbar^*]$ in the direction which further internalizes the externality. The default quality $q_d^*$ responds to the change in $\qubar^*$ or $\qbar^*$ following Result \ref{res:qd}, and the default transfer $c_d^*$ compensates this adjustment.
\end{result}

\subsection{Equal bargaining}\label{sec:app_equal}

We next turn to the case in which the worker and the firm have equal bargaining power, i.e. $\delta=\frac12$. This is a special case because the allocation of bargaining power is \textit{aligned} with the regulator's equity objective. More generally, we say that bargaining power is \emph{aligned} with social welfare when the regulator prefers to give the worker some $\alpha \in (0,1)$ share of the surplus, and the worker's bargaining power is $\delta=\alpha$. In such cases, bargaining will never exacerbate inequity. 

\subsubsection{Default delegation simplifies under aligned bargaining}

As a result of the alignment of bargaining power and social objectives, the optimal default delegation policy dramatically simplifies. The default $(q_d^*, c_d^*)$ plays a single role: it addresses the expected inequity between the parties. The delegation interval also plays a single role: it mitigates the externalities. We characterize the optimal default delegation policy in Proposition \ref{prop:aligned_bargaining}.

\begin{proposition}\label{prop:aligned_bargaining} \singlespacing 
Suppose the regulator's social welfare function is 
\begin{equation}\label{eq:swf_w_alpha}
  U_w + U_f-\beta (\alpha U_f -  (1-\alpha) U_w)^2 + \gamma U_r(q)  
\end{equation}
where \(\alpha\) represents the worker share of total surplus that is socially optimal. If \(\delta = \alpha\) then
\begin{itemize}
    \item[(i)] \(\underline{q}^*\) and  \(\overline{q}^*\) do not depend on \(q_d^*\) or \(c_d^*\), and
    \item[(ii)] \(q_d^*\) and \(c_d^*\) do not depend on \(\gamma\), \(\beta\),  \(\underline{q}^*\) or  \(\overline{q}^*\).
\end{itemize}
\end{proposition}
To illustrate this case more fully, we return to our example and solve for the optimal default-delegation policy under the uniform assumption (Assumption \ref{ass:uniform_unit}). The optimal default delegation policy is given by
\begin{equation}\label{eq:aligned_policy_unif}
q_d^* = \frac{1}{2} \quad c_d^* = \frac{R}{2} - \frac{1}{12} \quad  \underline{q}^* = \frac{\gamma}{2}.    
\end{equation}
The default quality $q_d^*$ is equal to the worker's ex-ante expected optimal quality level, \(\mu_\theta \equiv \int_{\underline{\theta}}^{\overline{\theta}}\theta dG(\theta).\) The regulator ``pays" the workers in kind by setting the default quality level $q_d^*$ to their expected optimal $\mu_\theta$. The default transfer $c_d^*$ instead favors the firm. This arrangement is optimal in this case because the workers have state-dependent preferences whereas the firm does not. The regulator minimizes the expected losses due to sub-optimal quality for the worker and equalizes the surplus through transfers to the firm. The minimum $\qubar^*$ is strictly a function of the externality term. Its slope is determined by equating the loss of efficiency to the gains coming from mitigating the externality.

\subsubsection{How changes in the regulator's prior affect default delegation}

We next take advantage of the simplicity of the aligned bargaining case to study how shifts in the regulator's prior influence the optimal default delegation policy. In particular, we compare the optimal default delegation policy  under the uniform distribution (Assumption \ref{ass:uniform_unit}), shown in \eqref{eq:aligned_policy_unif}, to the optimal default delegation policy under a distribution with higher density in higher states. 

\begin{assumption}\label{ass:dist_theta2}
Types $\theta$ are drawn from the distribution $G(\theta)= \theta^2.$
\end{assumption}
Under Assumption \ref{ass:dist_theta2}, the optimal default-delegation policy is given by
\begin{equation}\label{eq:aligned_policy_gtheta}
q_d^* = \frac{3}{5} \quad c_d^* = \frac{R}{2} -  \frac{3}{20} \quad \underline{q}^* = \frac{3\gamma }{4}. 
\end{equation}
Comparing \eqref{eq:aligned_policy_unif} and \eqref{eq:aligned_policy_gtheta} is instructive. As the cdf of $\theta$ shifts towards higher states, the default quality $q_d^*$ increases. However, the default quality level does not increase one for one with the expected state $\mu_\theta$. In the uniform case, \eqref{eq:aligned_policy_unif} shows that $q_d^* = \mu_\theta = \frac12$. With $G(\theta) = \theta^2,$ \eqref{eq:aligned_policy_gtheta} shows that $q_d^* = \frac35$ whereas \(\mu_\theta = \frac{2}{3}\). In the quadratic case, $\mu_\theta > q_d^*$ because the cost of raising the default quality $q_d^*$ is relatively high in the low states. Even so, the higher default quality level $q_d^*$ imposes a greater cost on the firm, which must be offset through the default transfer $c_d^*$: the transfer from the worker to the firm grows from $\frac{1}{12}$ in the uniform case to $\frac{3}{20}$ in the quadratic case. 

Meanwhile, the minimum $\qubar^*$ in the quadratic case is steeper in \(\gamma\) than in the uniform case. This is because the efficiency cost of raising the minimum is reduced due to the lower density in the low states. 

\begin{result}\label{res:prior_compstat}
Let $G'(\theta)$ be a distribution that first-order stochastically dominates $G(\theta)$. The optimal default quality level $q_d^*$ and the optimal minimum quality level $\qubar^*$ are higher under $G'(\theta)$ than under $G(\theta),$ and the optimal default transfer $c_d^*$ is lower. 
\end{result}


\subsection{Worker control}
Finally, we consider the case in which workers have all of the bargaining power (\(\delta = 1\)). Although this case does not align with the facts about platform work in any context we know of, it is a theoretically valuable case. In particular, it highlights the role of state-dependent preferences in the preceding analysis. 

\begin{proposition}\label{prop:state-independent bargaining}
Assume that only one agent $i$ has state-dependent preferences, and that this agent has all of the bargaining power (i.e. $\delta_i =1$). Then,
\begin{itemize}
    \item[(i)] the optimal default \(q_d^*, c_d^*\) is not uniquely determined, and
    \item[(ii)] the expected social welfare under the optimal default-delegation policy is strictly lower when $\delta_i=1$ than when $\delta_i = 0.$
\end{itemize}
\end{proposition}

The regulator has limited ability to distribute surplus across states when the agent with state-independent preferences has no bargaining power. Any attempt to improve equity necessitates a large efficiency loss. As a result, the expected welfare from the optimal default delegation policy is lower when the agent with state-independent preferences has no bargaining power. 


Proposition \ref{prop:state-independent bargaining} is related to Corollary \ref{cor: bargaining_and_default} in \autoref{sec:first_best}, which highlights that the set of implementable social choice functions is larger when the party that is less sensitive to the state has more bargaining power. In both cases, the regulator loses some of its ability to influence the state-dependent distribution of surplus when the party with more bargaining power is less sensitive to the state. In other words, when bargaining power shifts away from alignment and toward the agent with more sensitive preferences, the regulator's trade-off between discretion and control is exacerbated.

\section{Discussion}
 
 The model presented in this paper describes how a government optimally regulates contracting environments with \textit{default delegation} in order to maximize a general class of social welfare functions that incorporate efficiency, equity and externality concerns. Default delegation is an indirect mechanism in which the regulator's tools are limited: it can influence a specific default contract \(d\) and restrict the set of enforceable contracts \(\Q\). As illustrated with the examples in Section \ref{sec:examples}, default delegation captures key aspects of contract law (default and immutable rules) and social policy more broadly. 
 
 The intuition developed through the implementation-theoretic analysis in Section \ref{sec:first_best}, which focuses on a binary state case, lends insight into the primary roles of the default and the delegation set: the default is designed to achieve a particular distribution of surplus across states while the delegation set is constrained primarily to internalize externalities. Even when there is a continuum of states, as explored in Section \ref{sec:application}, Proposition \ref{prop:aligned_bargaining} shows that this simple logic behind default delegation is preserved when there is aligned bargaining. That is, when the regulator's distributive preferences align with the allocation of bargaining power, the default $(d=(q_d, c_d))$ controls the distribution of surplus while the delegation set $(\Q=[\qubar, \qbar])$ mitigates externalities. When bargaining is aligned, the expected social welfare under the optimal default delegation policy is closest to the first-best level of social welfare. 
 
 However, if the allocation of bargaining power is not aligned with the government's distributional preferences, the second-best optimal policy parameters become interdependent. This interdependence is shown most clearly in Result \ref{res:qd}, which shows that an increase in the optimal minimum employer-provided benefits (\(\underline{q}^*\)), all else equal, implies an increase in the optimal outside option benefits level (\(q_d^*\)).

Comparative statics on the second-best optimal default delegation policy offer potential explanations for recent attempts to raise minimum employer-sponsored benefits and platform workers' outside options. Three distinct shifts could lead the regulator to increase increase the minimum and the default quality level. First, when the externality parameter (\(\gamma\)) increases, as described in Result \ref{res:gamma_compstat}, the optimal default delegation policy increases the minimum quality level to mitigate the increase externalities, and, in so doing, increases the default quality level.

Second, when equity concerns (captured by the parameter $\beta$) increase, Result \ref{res:beta_compstat} shows that there is a further motive to constrain the delegation set and bring the default closer to the worker optimum. Indeed, Result \ref{res:beta_gamma_compstat} shows that as long as there are some externality concerns, any increase in equity concerns leads to an increase in the optimal minimum quality level. In such cases, the minimum quality level does ``double duty" in the sense that it mitigates externalities, and helps to achieve distributional objectives.

Third, when worker composition (captured by $G(\theta)$) changes such that more workers begin to value benefits higher relative to cash, Result \ref{res:prior_compstat} shows that the optimal default delegation policy sees an increase in the the optimal default and minimum benefits level.

Our analysis of second-best optimal default delegation helps to clarify the distinctive rationales behind increasing worker support \textit{in-kind} versus \textit{in cash}. Interestingly, all else equal, each of the three changes outlined above (which lead to higher benefits, i.e. ``in-kind") also imply a \textit{decrease} in the  default transfer $(c_d^*)$ (i.e. ``cash") in the optimal default delegation policy. Thus, calls for an increase in the minimum wage or unemployment insurance, captured by $c_d$, cannot be explained by changes in the governments' weighting of externalities or equity, nor by changes in worker composition starting from an initially optimal policy. However, calls for increases in minimum wages and unemployment could be justified either by an increase in the firm share of revenue $(R)$, or by political economy constraints that prevent the regulator from achieving the optimal default quality $(q_d^*)$.\footnote{As we've discussed, regulators may not be able to influence defaults---the workers' outside options---directly. Worker's outside options are influenced by labor laws but also by industry characteristics that may be outside the regulator's control. However, in practice a worker's outside option is an increasing function of the minimum level of benefits $\qubar$. The results discussed so far show that an increase in equity concerns in a context with externalities as is the case for app-based drivers would be sufficient for calls for increased minimum benefits.}

Calls for increases in minimum wages and unemployment insurance (i.e. $c_d$) can also be explained by a decline in worker bargaining power away from parity. A decline in worker bargaining power implies an increase in the optimal default transfer ($c_d^*$) to workers but a decrease in the default benefits level $q_d^*$.\footnote{Notice, by comparing \autoref{eq:aligned_policy_unif} and \autoref{eq:delta0_default_unconstrained}, the default quality level is actually higher when bargaining is aligned than when the worker has no bargaining power. This is because any gains from bargaining accrue to the firm. Although a decrease in bargaining power leads to an increase in minimum employer-sponsored benefits levels \(\underline{q}^*\), worker outside option benefits \(q_d^*\) may in fact decrease. Altogether, a change in bargaining power has a more nuanced impact on the optimal policy then simply shifting policy in favor of the worker.} The logic behind this result is that the regulator prefers to limit negotiation as worker bargaining power decreases---in this case, it is more efficient to equalize surplus across states through outside options' cash payments (unemployment insurance, minimum wage) than through benefits.



Beyond platform work, our analysis helps to untangle debates about how governments should design contracting environments to promote social welfare. In our model, contracting parties have symmetric unverifiable information and bargain efficiently and costlessly. Many regulatory arenas can be described by such assumptions---we discussed here examples from commercial, marriage, and employment contracts. When these assumptions hold, governments can influence parties' outside options to achieve particular distributional objectives, and they can limit the set of enforceable contracts in order to mitigate externalities.

\label{sec:conclusion}


\pagebreak 

\appendix
\section{Examples of Default Delegation in Contract Law and Social Policy} \label{sec:examples}
This section presents three contracting environments that the government influences through default delegation: commerce, marriage and employment. These examples show that concerns about efficiency, equity and externalities may show up under different guises. Further, these examples show that default delegation is sometimes carried out entirely through contract law---in the form of ``default" and ``immutable" rules (see \citet{AyresIan1989FGiI} for a discussion)---while other times default delegation is carried out through laws and policies that extend beyond contract law.

\paragraph{Example 1 (Commerce).}  \textit{The Uniform Commercial Code and ``Reasonable" Defaults}.\normalfont 

Commercial contracts in the United States are largely regulated by the Uniform Commercial Code (UCC), a law which standardizes contracting environments between commercial actors.\footnote{The UCC's primary objectives are to ``simplify, clarify and modernize the law governing commercial transactions; to permit continued expansion of commercial practices through custom, usage and agreement of the parties; to make uniform the law among the various jurisdictions."} One of the key features of this standardization is to make clear how courts will enforce contracts. There are two aspects of enforcement that align with the features of default delegation: (i) a statement of \textit{immutable rules}, i.e. contract terms which courts \textit{will not} enforce (``delegation set") and (ii) a statement of \textit{default rules}, i.e. terms the court will enforce if the parties do not specify otherwise (``default outcome"). 

There is relative consensus in the legal literature about the role of immutable rules: such rules internalize externalities. Meanwhile, there has been significant debate with regards to the role of default rules. Our model highlights that default rules allow the regulator to achieve a desired distribution of surplus across unverifiable states. For instance, as argued by \cite{easterbrook1989corporate}, the default can be used in the context of the UCC codes to fill in for what the contracting parties ``would have wanted" by providing an ex ante desirable distribution of surplus. 

The defaults in the UCC hold when the contracting parties do not explicitly state an alternative. For instance, the UCC states that ``time for shipment or delivery... if not provided in this article or agreed upon shall be a reasonable time." The ``reasonable time" can be chosen in order to solve an incomplete contracting problem in a manner similar to \cite{aghiondewatripontrey1994}, where contracting parties engage in ``renegotiation design" by committing to a default rule ex-ante to be renegotiated ex-post. A distinction between the UCC defaults and ``renegotiation design" defaults is that the government's position outside of the contract enables it to commit to a particular ``reasonable" default after the contract has been signed. This means that the courts can fill in their definition of ``reasonableness" with new verifiable information has been revealed about the true state of the world. The government's commitment to enforce a reasonable default provides an incentive for contracting parties to omit particular aspects of a contract, in anticipation of potentially non-contractible but verifiable information revelation.

For instance, a supplier with a long term contract with a buyer may not want to stipulate a standard time for delivery if there may be a future state of the world which will make it prohibitively costly to follow the original contract. Note that, in this case, renegotiation from the original contract may lead to a much more costly outcome for the supplier relative to the contract the buyer and supplier would have agreed to were they able to contract on the unforeseen contingency.  

Our model is consistent with an optimal policy of enforcing a ``reasonable" default which achieves a distribution of surplus consistent with what the contracting parties would have chosen if they had contracted on the unforeseen event. 

\paragraph{Example 2 (Marriage).}  \textit{The ``Equitable Division" Default and Limited Enforcement of Premarital Contracts}.

\normalfont
Another setting which has fallen under the purview of a uniform act in the U.S. is the law governing premarital agreements. The Uniform Premarital Agreement Act (UPAA), adopted (often with slight differences) by 28 states, specifies the default marriage contract which holds in the absence of an alternative premarital agreement. Beyond the default rules outlined in the UPAA, states can also decide the extent to which they will enforce particular alternative contracts. Just like the UCC for commercial contracts, the UPAA influences the design of marriage contracts through ``default delegation." 

In the case of marriage, the government explicitly considers issue of fairness when choosing default and immutable rules \citep{BixBrian1998Bits}. When it comes to the division of assets in the event of a divorce laws, the majority of states enforce an ``equitable division" default rule \citep{HerschJoni2019Wein}.\footnote{``Equitable division rules often end up splitting assets 50-50 in terms of monetary value, but legally take into account a range of factors to determine what is ``equitable." For example, in Mass. Gen. Laws ch. 208 § 34 (2018), the division of assets upon divorce is decided through a holistic evaluation of ``the length of the marriage, the conduct of the parties during the marriage, the age, health, station, occupation, amount and sources of income, vocational skills, employability, estate, liabilities and needs of each of the parties."} However, to-be-spouses can contract around ``equitable division" default rules by explicitly specifying how assets are to be divided in the event of divorce. The Uniform Premarital Agreement Act puts few limitations on pre-martial contracts, with the primary exceptions being: (i) the agreed division of assets cannot be ``extremely unfair" and involve a lack of disclosure, and (ii) the agreed division of assets cannot necessitate government assistance for one of the spouses. 

Courts will also not enforce any premarital contract terms related to the division of custody. These limits on premarital contracts are immutable laws that specifically address the presence of externalities which affect the government and potential children. 

In the previous example, the UCC supplies a ``reasonable" default delivery time when not otherwise specified, at least in part to ensure particular distributions of surplus in ex-ante noncontractible states of the world. In marriage contracts, the ``equitable division" default rules serve a similar purpose, however, in this case, the government has explicit distributional concerns for fairness reasons. 


\paragraph{Example 3 (Employment).}  \textit{The Classification of Platform Workers.}\normalfont

The last two examples lean on the presence of (or potential for) incomplete contracts as a reason for government intervention in the form of default and immutable rules. This paper argues that the regulation of bilateral contracts more generally can be viewed through the lens of default delegation: default rules affect the agents' disagreement outcome for bargaining in order to affect the distribution of surplus across unverifiable states of the world, while limits on enforceable contracts curb externalities.

To show the logic of default delegation, we will focus on the example of regulating app-based platform workers, which has led to contentious debate in the U.S. and abroad. The debate has largely focused on whether these workers should be classified as employees or independent contractors. As independent contractors, there are very few limitations on the contracting space. On the other-hand, employee status entails a minimum suite of benefits as well as firm contributions for programs such as social security, unemployment insurance, payroll taxes, and premiums for worker's compensation. 

Thus, the imposition of employee status, constrains the set of contracts that the government enforces. More subtly, the change in employment status also affects the workers' outside options, and thus their disagreement outcome in employment negotiations. Employee status raises the minimum levels of benefits and wages that workers' can expect to receive from a different company. In this sense the government can indirectly influence a worker's outside option.

The recent debate around the classification of app-based platform workers has centered around the distinctive nature of this new form of work. California's Assembly Bill No. 5 (AB5) attemped to clarify the employment status of app-based platform workers in order ``to ensure workers who are currently exploited by being misclassified as independent contractors instead of recognized as employees have the basic rights and protections they deserve under the law, including a minimum wage, workers’ compensation if they are injured on the job, unemployment insurance, paid sick leave, and paid family leave." In the context of this paper, this can be seen an attempt to increase the default through raising minimum benefits in response to perceived inequity in these contracting arrangements. 

After the passage of AB5, Lyft and Uber spent more than \$200MM in order to pass a referendum, Proposition 22, which returned the employment classification of app-based drivers to independent contractors. Proposition 22 did however include some minimum earnings requirements, non-discrimination protections and some minimum health insurance.\footnote{\url{https://pmcinsurance.com/blog/how-assembly-bill-5-differs-from-proposition-22/}} In effect, Proposition 22 created a distinct classification due to the unusual contracting environment surrounding app-based drivers, which could neither be viewed simply as employees or as traditional independent contractors.  

The rhetoric surrounding the debates over Proposition 22 were largely focused on questions of equity versus efficiency.\footnote{See, for example,

\url{https://www.americanactionforum.org/insight/whats-next-for-prop-22-and-debates-around-independent-workers/}} Those that opposed Proposition 22 argued that the classification of workers as employees would improve their expected outcomes by entitling them to benefits and wage standards. Classifying workers as employees would lead to a more equitable split of the total surplus generated from these contracts. The proponents of Proposition 22 argued that platform workers have a relatively low valuation for employment benefits that come with employee status and profit from being able to accept contracts with lower benefits and higher pay. The model below captures the key tradeoffs of regulation in this environment. In \autoref{sec:application}, we return to the regulation of platform work in depth, fleshing out how features of the contracting environment and parameters in the relevant social welfare function dictate optimal regulatory policy.  

\section{Three State Case}
\label{sec:unconstrained}
\label{sec:threestates}



We first generalize Proposition \ref{prop:fb_with_gamma_beta}  to more than two states in Proposition \ref{prop:multiple_states} below. Then we discuss the implications. 

\begin{proposition}\label{prop:multiple_states}
Assume $2<|\Theta|<\infty$. Then a social welfare function $\SWF$ is DD-implementable if
\begin{itemize}
    \item[(i)] there exists a $d=(q_d, c_d)$ satisfies
    \begin{equation}
        (1-\delta)[\Delta\Delta(u_w, q_d, \theta, \theta')] - \delta[\Delta\Delta(u_f, q_d, \theta, \theta')] = c_{\theta'} - c_\theta 
    \end{equation}
    for all $\theta, \theta'$, and
    \item[(ii)] $\Q = \{q_\theta\}_{\theta \in \Theta}$ satisfies \eqref{eq:fb_with_gamma}.
\end{itemize}
\end{proposition}
The first hypothesis (i) in Proposition \ref{prop:multiple_states} (about general $\SWF$s) contrasts with the result of Proposition \ref{prop:imp_default} (about general $\betaswf$s). Default delegation \emph{does not} implement the full set of implementable SWFs when there are more than two states, even when $\gamma=0$ and $U_f$ and $U_w$ are continuous in $q$ and $\theta$.

To give intuition for why this is the case,  we consider a setting with three states, i.e. when the state space is $\Theta = \{\theta_l, \theta_m, \theta_h\}$. 

As before, it is useful to study the direct mechanism in which the regulator is restricted to using message-independent threats. In this case the game takes the form presented in \autoref{tab:default_game_3states} (which is a simple extension of \autoref{tab:default_game_2states} to three states). 

    \begin{table}[H]
     \centering
    \begin{tabular}{c|c|c|c}
            & \(\thetahat_f = \theta_l\) & \(\thetahat_f = \theta_m\) & \(\thetahat_f = \theta_h\)\\
            \hline  & & \vspace{-0.35cm} \\  
                \(\thetahat_w = \theta_l\) & \((q_l, c_l)\) & \((q_d,c_d)\) & \((q_d,c_d)\) \\
                 \hline 
                  & & \vspace{-0.4cm} \\  
                \(\thetahat_w = \theta_m\) & \((q_d,c_d)\) & \((q_m,c_m)\) &  \((q_d,c_d)\)\\
                 \hline 
                  & & \vspace{-0.4cm}\\
                \(\thetahat_w = \theta_h\) & \((q_d,c_d)\) & \((q_d,c_d)\) &  \((q_h,c_h)\)\\
                 \hline 
            \end{tabular}
     \caption{Direct mechanism with message-independent threats $(|\Theta|=3)$.}
     \label{tab:default_game_3states}
 \end{table}
 In general, it will not be possible to find a \((q_d,c_d)\) that implements the first-best outcomes. This is due to the constraints imposed by equation \eqref{eq:default_cond}, which relates the differences in utility at $q_\theta$ and $q_d$ in each state to the difference in desired transfers $c_h-c_l$. When there is a third state, there is now another value $c_m$, which appears in two other conditions analogous to \eqref{eq:default_cond}. In order to implement the first-best, the value $c_m$ must satisfy  
\begin{equation}\label{eq:13_multi_state}(1-\delta)[\Delta\Delta(u_w,q_{d},\theta_h, \theta_m)] - \delta[\Delta\Delta(u_f,q_{d}, \theta_h, \theta_m)] = c_m - c_h\end{equation}
\begin{equation}\label{eq:23_multi_state}(1-\delta)[\Delta \Delta (u_w,q_{d},\theta_m, \theta_l)] - \delta[\Delta \Delta(u_f,q_{d}, \theta_m, \theta_l)] = c_l - c_m\end{equation}
where again $\Delta\Delta(u, q, \theta, \theta') \equiv \Delta(u,q,\theta)-\Delta(u,q, \theta').$ In general, it will not be the case that $c_m$ satisfies both of these equations. So, there are outcomes $(q^*, c^*)$ that are implementable with a general direct mechanism that are not implementable when the regulator is restricted to message-independent threats. Thus, Proposition \ref{prop:imp_default} does not hold when there are three states. In fact, as the following proposition highlights, Proposition \ref{prop:imp_default} does not hold when $|\Theta|>2.$

\section{Interpretation of First-Best Results in the Commercial, Marriage and Employment Contracts}

We interpret the results of the first-best analysis in the context of the examples presented in \autoref{sec:examples}. \autoref{tab:interpretation} outlines examples of defaults and limits in commercial law, marriage law and labor law. 

\begin{table}[h]
    \centering
    \footnotesize
    \begin{tabular}{|c|c|c|c|} 
    \hline
          \# & Example & Default $d$ & Limits $\mathcal Q$  \\
         \hline \hline 
         \textbf{1} & \textbf{Commerce} & ``reasonable" times & no ``manifestly unreasonable" delivery times \\
         \hline
            \textbf{2} & \textbf{Marriage} & ``equitable division" of assets & no terms about custody \\
                \hline
            \textbf{3} & \textbf{Work} & ``employee" classification & minimum health insurance coverage \\ \hline 
    \end{tabular}
    \caption{Examples of Defaults and Limits in Settings from Section \ref{sec:examples}}
    \label{tab:interpretation}
\end{table}

\paragraph{Example 1 (Commerce).} \textit{The Uniform Commercial Code and ``Reasonable" Defaults.}

We can use the results of this section to understand aspects of the Uniform Commercial Code. Consider first the issue of contracting on delivery times. Suppose that $\theta$ takes on only two states, leading to a high or a low surplus, respectively. Section \ref{sec:examples} showed that without a default rule, the ``low" realization of $\theta$ may be so bad for the seller that the seller would avoid entering the contract to begin with. In other words, the distribution of surplus in this state is such that the seller would not enter. In this scenario, the government has an ``equity" objective ($\beta>0$), to smooth the distribution of surplus, which is based in a goal of providing efficient investment incentives.  Proposition \ref{prop:imp_default} suggests that when the government has some ``equity" objective $\beta>0$, it can implement a ``reasonable" default rule for delivery times which achieves the first best distribution of surplus (and therefore the efficient contract) in both states of the world \(\{\theta_l,\theta_h\}\). 

The U.C.C. also has some immutable rules, which place limits on the kinds of contracts that will be enforced in court. For instance, the law prevents contract terms stipulating times deemed ``manifestly unreasonable." If times for delivery are too extreme, there will likely be the need for arbitration. Arbitration can be costly and therefore strain public resources. In other words, there is an externality that arises out of contracts with extreme times for delivery, which is mitigated with the ``manifestly unreasonable" immutable rule. 

\paragraph{Example 2 (Marriage).} \textit{``Equitable Division" Defaults and Limited Enforcement.}

In the regulation of marriage contracts in the U.S., the government supplies default marriage rules that govern unless expressly contracted around. For instance, if to-be-spouses get married without signing a prenuptial agreement that specifies otherwise, the division of assets upon divorce will follow an ``equitable division" rule in many states. The default rule fills a gap in an incomplete contract, and ensures a particular distribution of surplus in the event of divorce. In this case, unlike in Example 1, the government may have preferences directly over the ``equity" of a contract, for fairness reasons \citep{BixBrian1998Bits}. That is $\beta>0$ because the government wants to avoid unjust outcomes. Proposition 1 thus helps to illustrate why this equitable division rule exists: it serves as a default that allows the regulator to achieve first best. 

Proposition 3 can help to understand the limits on the contracts the government is willing to enforce. The best allocation of custody from the perspective of the government is the allocation that is best for the child. This allocation may not align with what the spouses view to be the best allocation. Suppose the government has an infinitely negative payoff when full custody is given to the ``wrong" spouse. Then, our model predicts that the government would not enforce \textit{any} contract terms involving custody of children, child support, or visitation. In line with this prediction of our model, there is no state in the U.S. which will enforce terms about children in prenuptial contracts.\footnote{\href{https://www.findlaw.com/family/marriage/what-can-and-cannot-be-included-in-prenuptial-agreements.html}{https://www.findlaw.com/family/marriage/what-can-and-cannot-be-included}}



\paragraph{Example 3 (Work).} \textit{The Classification of Platform Workers}.

We return to the case of regulating app-based platform work in detail in section \ref{sec:application}. For now, it is sufficient to establish the premise that to the extent that the regulator is able to affect the default, they will choose to shift the default contract in order to push the resulting distribution of outcomes across states closer to their preferred distribution. Furthermore, the presence of externalities justifies the restriction of the contracting space inherent in mandating minimum health and unemployment insurance coverage. 

One aspect of this setting that makes interpretation more subtle is that there are multiple policy levers through which the government influences ``default" labor contracts, i.e. the labor contracts that serve as an outside option in specific employee-employer negotiations. (This subtlety contrasts with the incomplete contracting setting where the regulator directly establishes defaults which hold in case they are not explicitly contracted away.) A primary avenue through which governments affect worker ``defaults" is worker classifications, which establish that certain forms of work in certain sectors must be governed by ``employee" rules and not ``independent contractor" rules.  Regulators also use the threat of legislation and other forms of soft power to shift the default \citep{StewartAndrew2017Rwit}. For instance, Bernie Sanders' Stop BEZOS Act may be partially responsible for Amazon's subsequent decisions to raise wages.\footnote{See, for example, \citet{wapostopbezos}.} 

In the platform work example, the unverifiable state of the world $\theta$ is the degree to which workers value benefits relative to cash. Direct surveys show that there is a large degree of heterogeneity in $\theta$ (e.g. \citet{NBERw29736}) justifying our continuous treatment of \(\Theta\). We use the platform work example to characterize the optimal default delegation policy. Comparative statics on the optimal policy help us understand recent and anticipated shifts in the approach to regulating platform work.

\section{Proofs}\label{sec:proofs}

\subsection*{Proposition \ref{prop:maskinmoore}.}

See \citet{maskinmoore1999}, Theorem 2.

\subsection*{Proposition \ref{prop:imp_default}.}
We take as a starting point the reduced form game shown in \autoref{tab:general_game_2states}:

\begin{table}[H]
     \centering
        \begin{tabular}{c|c|c}
            & \(\thetahat_f=\theta_l\) & \(\thetahat_f=\theta_h\)   \\
            \hline 
            & & \vspace{-0.35cm} \\   
                \(\thetahat_w=\theta_l\) & \((q_l, c_l)\) & \((q_{lh},c_{lh})\) \\
                 \hline 
                 & & \vspace{-0.4cm} \\   
                 \noalign{\vskip 1mm}    
                \(\thetahat_w=\theta_h\) & \((q_{hl},c_{hl})\) & \((q_h,c_h)\) \\
                 \hline 
            \end{tabular}
     \caption{Direct mechanism for implementing $(q_\theta, c_\theta)$ in state $\theta$.}
\label{tab:general_game_2states}
 \end{table}

Utilities from strategy \(\theta_h, \theta_l\) will depend on who deviated. Specifically, if \(w\) deviates in state \(\theta_l\), their payoff will be 

\[u_w(q_{hl}; \theta_l)+\delta \left(\sum_i u_i(q_l,\theta_l) - \sum_i u_i(q_{hl},\theta_l)\right)+c_{hl}\]
If \(f\) deviated in state \(\theta_h\), their payoff will be 
\[u_f(q_{hl}; \theta_h)+(1-\delta)\left(\sum_i u_i(q_h,\theta_h) - \sum_i u_i(q_{hl},\theta_h)\right)-c_{hl}\]
In order for \(w\) and \(f\) to not deviate in these cases, the following two inequalities must hold: 
\begin{align}
\label{eq:chl_1}
    (1-\delta)\left(u_w(q_{hl},\theta_l) -  u_w(q_l,\theta_l)\right)-\delta\left(u_f({hl},\theta_l) -  u_f(q_l,\theta_l)\right) &\leq c_l - c_{hl} \\ \label{eq:chl_2}
    \delta\left(u_f(q_{hl},\theta_h) -  u_f(q_h,\theta_h)\right)-(1-\delta)\left(u_w(q_{hl},\theta_h) -  u_w(q_h,\theta_h)\right) &\leq c_{hl} - c_h
\end{align}
Define \(\underline{c_{hl}}\) to be the smallest value of \(c_{hl}\) such that \eqref{eq:chl_1} holds. That is,
\[ \underline{c_{hl}} = c_l -   (1-\delta)\left(u_w(q_{hl},\theta_l) -  u_w(q_l,\theta_l)\right)+\delta\left(u_f(q_{hl},\theta_l) -  u_f(q_l,\theta_l)\right).  \]
We can then plug this expression into equation \ref{eq:chl_2} to get  
\begin{equation}\label{eq:ineq_1}
 (1-\delta)[\Delta(u_w, q_{hl}, \theta_h) -  \Delta(u_w, q_{hl}, \theta_l)] - \delta[\Delta(u_f, q_{hl}, \theta_h) -  \Delta(u_f, q_{hl}, \theta_l)] \geq c_l - c_h
\end{equation}
where $\Delta(u, q,\theta) \equiv u(q_\theta, \theta) - u(q, \theta).$ We can do a similar calculation for the other off diagonal entry resulting in: 
\begin{equation}\label{eq:ineq_2}
 (1-\delta)[\Delta(u_w, q_{lh}, \theta_h) -  \Delta(u_w, q_{lh}, \theta_l)] - \delta[\Delta(u_f, q_{lh}, \theta_h) -  \Delta(u_f, q_{lh}, \theta_l)] \leq c_l - c_h
\end{equation}
This shows that in order for default delegation to implement the SWF there must exist a \(q_d\) such that \(q_d = q_{hl} = q_{lh}\) will satisfy both both \eqref{eq:ineq_1} and \eqref{eq:ineq_2} with equality.

Now we want to show that any implementable mechanism can be implemented by default delegation. It is sufficient to show that any mechanism which implements a specific difference in transfer \(c_l-c_h\), can be implemented by choosing \(q_d = q_{hl} = q_{lh}\). Beginning from Proposition \ref{prop:maskinmoore}, implementability requires finding \(q_{lh}\) and \(q_{hl}\) which satisfy equations \ref{eq:ineq_1} and \ref{eq:ineq_2}. If \(u_w\) and \(u_f\) are smooth, then it must be the case that if there exist \(q_{lh}\) and \(q_{hl}\) which satisfy equations \ref{eq:ineq_1} and \ref{eq:ineq_2}, then there must exist \(q \in [q_{lh},q_{hl}]\), which satisfy them with equality. This is the default in the optimal default delegation mechanism. 

\subsection*{Corollary 1.}
We have shown that when \(u_w\) and \(u_f\) are continuous, if the first best is implementable, then there exists \(q_d\) such that 

        \begin{align}\label{eq:cor_proof_1}
             (1-\delta)&  [(u_w(q_{h},\theta_h) - u_w(q_d, \theta_h)) - (u_w(q_{l},\theta_l) - u_w(q_d, \theta_l))] \\ \nonumber
         -\delta & [(u_f(q_{h},\theta_h) - u_f(q_d, \theta_h))-(u_f(q_{l},\theta_l) - u_f(q_d, \theta_l))] \\ \nonumber
             = & c_l - c_h 
        \end{align}
where we've expanded terms in equation \eqref{eq:default_cond}. We begin the proof by assuming that the condition from the corollary is met such that
\[\left|(1-\delta)\frac{\partial^2 u_w(q,\theta)}{\partial q \partial \theta} - \delta \frac{\partial^2 u_f(q,\theta)}{\partial q \partial \theta}\right| > x\]
where \(x \in \mathbb{R}_{+}\). Furthermore, we will prove this for the specific case where
\begin{equation}\label{eq:cor_proof_2}
    (1-\delta)\frac{\partial^2 u_w(q,\theta)}{\partial q \partial \theta} - \delta \frac{\partial^2 u_f(q,\theta)}{\partial q \partial \theta} > x
\end{equation}
noting that the symmetric argument will hold in the opposite case. We can rewrite equation \eqref{eq:cor_proof_1}, above as 

\begin{align*}\label{eq:int_constraint12}
      (1-\delta)\int_{\theta_l}^{\theta_h} \int_{q_d}^{q_{h}} \frac{\partial^2 u_w(q,\theta)}{\partial q \partial \theta} dqd\theta - \delta\int_{\theta_l}^{\theta_h} \int_{q_d}^{q_{h}} \frac{\partial^2 u_f(q,\theta)}{\partial q \partial \theta} dqd\theta \\
      +(1-\delta)\left( u_w(q_h,\theta_l)-u_w(q_l,\theta_l)\right) - \delta\left( u_f(q_h,\theta_l)-u_f(q_l,\theta_l)\right)\\& = c_l - c_h
 \end{align*}
  where we have used the fundamental theorem of calculus to substitute
\[ \int_{\theta_l}^{\theta_h} \int_{q_d}^{q_h} \frac{\partial^2 u_i(q,\theta)}{\partial q \partial \theta} dqd\theta = u_i(q_{h},\theta_h)-u_i(q_{d},\theta_h)  -\left(u_i(q_{h},\theta_l)-u_i(q_{d},\theta_l) \right).
\]
Define \((1-\delta)\left( u_w(q_h,\theta_l)-u_w(q_l,\theta_l)\right) - \delta\left( u_f(q_h,\theta_l)-u_f(q_l,\theta_l)\right)=\kappa\), with $\kappa \in \mathbb{N}_+$. An outcome is implementable if there exists \(q_d\) which satisfies

\begin{equation}\label{eq:int_imp}
    (1-\delta)\int_{\theta_l}^{\theta_h} \int_{q_d}^{q_{h}} \frac{\partial^2 u_w(q,\theta)}{\partial q \partial \theta} dqd\theta - \delta\int_{\theta_l}^{\theta_h} \int_{q_d}^{q_{h}} \frac{\partial^2 u_f(q,\theta)}{\partial q \partial \theta} dqd\theta = c_l - c_h -\kappa 
\end{equation}

Now using our assumption \eqref{eq:cor_proof_2} we know that 
\[(1-\delta)\frac{\partial^2 u_w(q,\theta)}{\partial q \partial \theta}-\delta \frac{\partial^2 u_f(q,\theta)}{\partial q \partial \theta} dqd\theta > x\]
This means that beginning from \(q_d=q_h\) and decreasing the default we can achieve any positive value of \(c_l-c_h-\kappa\). By increasing the default we can achieve any negative value of \(c_l-c_h-\kappa\). Thus, we have proven that when equation \ref{eq:cor_proof_2} holds and \(\gamma =0\) such that the first best aligns with the agent negotiated outcome, the first best is implementable. The same argument holds changing the inequality in equation \ref{eq:cor_proof_2}.

\subsection*{Corollary 2.}
The result stems directly from the proof for Corollary 1. Note that as long as the condition in corollary 1 is met such that 
\[\frac{\partial^2 U_w}{\partial q \partial \theta} = b > 0 \hspace{3mm}\text{  and   }\hspace{3mm} \frac{\partial^2 U_f}{\partial q \partial \theta} = -a < 0\]
with \(b>a\), then decreasing \(\delta\) will increase

\[(1-\delta)\frac{\partial^2 u_w(q,\theta)}{\partial q \partial \theta}-\delta \frac{\partial^2 u_f(q,\theta)}{\partial q \partial \theta} dqd\theta\]
Note that this implies that for \(q_d < q_h\), the left hand side of \autoref{eq:int_imp} is decreasing in \(\delta\). For \(q_d > q_h\), it is decreasing in \(\delta\). Meanwhile, the left hand side of \autoref{eq:int_imp} is strictly increasing in \(q_d\). Therefore, a given value of \(c_l-c_h-\kappa\) and thus, a given transfer, can be implemented with a level of \(q_d\) closer to \(q_h\) if \(\delta\) is lower. In order to implement \(c_l-c_h-\kappa<0\), \(q_d > q_h\). Since lowering \(\delta\) decreases the left hand side of \autoref{eq:int_imp}, we have that the same transfer can be achieved with a smaller \(q_d\) (closer to \(q_h\)). A similar argument suffices for the case when \(c_l-c_h-\kappa>0\).

\subsection*{Proposition \ref{prop:fb_with_gamma}}

We continue to assume that agents Nash bargain and therefore the agents choose \((q_\theta,c_\theta)\) satisfying: 
\begin{equation}\label{eq:prop3_max}
\max_{q_\theta \in \Q, c_\theta \in \mathbb{R}_{+}} ( U_w(q_\theta,c_\theta;\theta) - U_w(d;\theta))^\delta( U_f(q_\theta,c_\theta;\theta) - U_f(d;\theta))^{1-\delta}
\end{equation}
Note that because the transfers are unrestricted,  bargaining will be constrained efficient. The agents will choose \(q_\theta\) to maximize total surplus and then choose the transfers in order to satisfy \eqref{eq:prop3_max}. 

Another way to see this is through a contradiction. Assume that agents have chosen \(q_{\theta'}\) in state \(\theta\). This implies:

\begin{align*}( U_w(q_{\theta'},c_{\theta'};\theta) - U_w(d;\theta))^\delta( U_f(q_{\theta'},c_{\theta'};\theta) - U_f(d;\theta))^{1-\delta} \\ > ( U_w(q_{\theta},c_{\theta};\theta) - U_w(d;\theta))^\delta( U_f(q_{\theta},c_{\theta};\theta) - U_f(d;\theta))^{1-\delta}\tag{(*)}\end{align*}
Next, posit that 
\begin{equation}\label{eq:jointsurplus}
    u_w(q_{\theta},\theta)+u_f(q_{\theta},\theta) > u_w(q_{\theta'},\theta)+u_f(q_{\theta'},\theta),
\end{equation}
i.e. the agents' joint surplus from $q_\theta$ in state $\theta$ is higher than their joint surplus from $q_{\theta'}$ in state $\theta.$

We can show that if \eqref{eq:jointsurplus} holds, then \(q_{\theta'}\) cannot satisfy ($*$) and therefore would not satisfy equation \eqref{eq:prop3_max}. To see this, note that if \eqref{eq:jointsurplus} holds then there must be a transfer \(c_\theta\), which satisfies

\[U_f(q_\theta,c_\theta;\theta) - U_f(d;\theta) \geq U_f(q_{\theta'},c_{\theta'};\theta) - U_f(d;\theta)\]
and 
\[U_w(q_{\theta},c_{\theta};\theta) - U_w(d;\theta) \geq U_w(q_{\theta'},c_{\theta'};\theta) - U_w(d;\theta).\]
Furthermore, one of the two inequalities must be strict, and so we have a contradiction to ($*$) and therefore \eqref{eq:prop3_max}. The agents would not choose \(q_{\theta'}\) in state $\theta$ as long as \eqref{eq:jointsurplus} holds. 

Note that this is simply an outcome of bargaining being constrained efficient. It is also useful to note that because in this case we have \(\beta = 0\) the default does not play a role in achieving the first best as the regulator is unconcerned about the resulting transfers.

\subsection*{Proposition \ref{prop:fb_with_gamma_beta}}

From the proof of Proposition \ref{prop:fb_with_gamma}, the agents will only choose the corresponding quality if the corresponding conditions are satisfied. Note that if this is the case then we transpose into the identical situation as Proposition \ref{prop:imp_default}. Therefore, the same argument applies. 

\subsection*{Proposition \ref{prop:aligned_bargaining}} 
This proposition is derived in Appendix Section \ref{sec:App_der} 5.2.

\subsection*{Proposition \ref{prop:state-independent bargaining}} 
This proposition is derived in Appendix Section \ref{sec:App_der} 5.3.

\subsection*{Proposition \ref{prop:multiple_states}}
This a mutli-state extension of Proposition \ref{prop:fb_with_gamma_beta} and the proof follows directly.

\section{Max-Min Social Welfare Functions} \label{sec:maxmin}

We show that when a regulator has a maxmin objective function, the results characterizing first-best implementability with default delegation apply when there are more than two states. 

A max-min objective function takes the form 
\begin{equation}
    \max_{\Q, d}\min_{\theta}\E[\SWF(q^*, c^*; \theta)]
\end{equation}
subject to incentive constraints. We say a SWF is \emph{max-min implementable} if the first-best outcome corresponding to the worst-case state can be implemented in Nash equilibrium and any refinement.

Suppose $\theta_k$ is the state that delivers the worst welfare for the principal, i.e.
$$\theta_k \in \arg\min_\theta \SWF(q,c;\theta).$$
Then, the principal can partition the state space $\Theta$ into cells $\{\theta_1, \theta_2, \dots \},\{\theta_k\}$. The states that are not $\theta_k$ can be treated as one, and the principal can offer a delegation set $\{(q_{-k},c_{-k}), (q_k, c_k)\}$ and a use a default that satisfies \eqref{eq:default_cond}. Although social welfare will not be ``first-best" in all states that are not $k$, it can be first best in state $k$.

 Consider the following example: 

\begin{example}\label{ex:gamma_not0}
Consider the following specification of preferences:
\begin{itemize*}
\item $u_w(q, \theta) = -(q-\theta)^2$ (worker's bliss point $q= \theta$), 
\item $u_f(q) = -q^2$ (firm's bliss point $q=0$), and
    \item $\Theta = \{\theta_l, \theta_m, \theta_h\} = \{0, \frac12, 1\}.$
\end{itemize*}
With these preferences, the regulator's welfare is minimized in the high state $\theta_h$. So, a max-min regulator cares only about maximizing welfare in state $\theta_h.$ In such a case, the regulator can design the direct mechanism in \autoref{tab:default_game_maxmin_3states}. The first-best outcome in the high state is $(q_h, c_h)=(\frac{\theta}{2}, \frac{1}{4})$.
\begin{table}[H]
     \centering
    \begin{tabular}{c|c|c}
            & \(\thetahat_f \in \{\theta_l, \theta_m\}\) &\(\thetahat_f = \theta_h\)\\
            \hline  & & \vspace{-0.35cm} \\  
                \(\thetahat_w \in \{\theta_l, \theta_m\}\) & \((q_{-h}, c_{-h})\) & \((q_d,c_d)\) \\
                 \hline 
                  & & \vspace{-0.4cm} \\  
                \(\thetahat_w = \theta_h\) & \((q_d,c_d)\) &  \((q_h,c_h)\)\\
                 \hline 
            \end{tabular}
     \caption{Default delegation with max-min regulator $(|\Theta|=3)$.}
     \label{tab:default_game_maxmin_3states}
 \end{table}
The conditions for implementation of $(q_{-h}, c_{-h})$ in states $\theta_l$ and $\theta_m$ and $(q_{h}, c_h)$ in state $\theta_h$ are 
\begin{equation}
    (1-\delta)[\Delta(u_w, q_{h}, q_d, \theta_h)-\Delta(u_w, q_{-h}, q_d, \theta_{-h})] -\delta[\Delta(u_f, q_{h}, q_d, \theta_h)-\Delta(u_f, q_{-h}, q_d, \theta_{-h})] \geq c_{-h}-c_h
\end{equation}
and 
\begin{equation}
    (1-\delta)[\Delta(u_w, q_{h}, q_d, \theta_h)-\Delta(u_w, q_{-h}, q_d, \theta_{-h})] -\delta[\Delta(u_f, q_{h}, q_d, \theta_h)-\Delta(u_f, q_{-h}, q_d, \theta_{-h})] \leq c_{-h}-c_h
\end{equation}
for $\theta_{-h} \in \{\theta_l, \theta_m\}.$ The regulator here has many degrees of freedom to choose $(q_{-h}, c_{-h})$ and $(q_{d}, c_{d})$ so that the constraints above are both satisfied with equality for  $\theta_{-h} \in \{\theta_l, \theta_m\}$, i.e. 
\begin{equation}
    (1-\delta)[\Delta(u_w, q_{h}, q_d, \theta_h)-\Delta(u_w, q_{-h}, q_d, \theta_{-h})] -\delta[\Delta(u_f, q_{h}, q_d, \theta_h)-\Delta(u_f, q_{-h}, q_d, \theta_{-h})] = c_{-h}-c_h.
\end{equation}

\end{example}
This example shows that even with $|\Theta|>2,$ the two state case offers valuable insight. For instance, when a regulator has max-min preferences, the results from the two state case (under some further conditions) show us that the regulator can implement the first best in the worst-case state. 

More generally, we can learn from the two state cases in order to understand the relative importance of particular states. Note that in the case where the regulator only has concerns about efficiency and equity, the optimal transfers lead to equal surplus in a particular state. Thus, the deviation from optimal can be measured by the difference between the optimal transfer and the transfer induced by a particular default. Thus, we can imagine solving for \(q_d, c_d\) by using \autoref{eq:13_multi_state} and then plugging in that value into equation \autoref{eq:23_multi_state} in order to solve for the transfer outcome \(\tilde{c}_2\) where the tilde denotes that it is not necessarily optimal. You could similarly perform this exercise to extract \(\tilde{c}_1\). The losses from optimizing across states 1 and 3 would be smaller than the losses from optimizing across 2 and 3 if \(|\tilde{c}_2-c_2|<|\tilde{c}_1-c_1|\).

\section{Extensions}

\subsection{Inequity penalty in SWF.}\label{sec:inequity_penalty}  Our model assumes that the social welfare function the regulator wants to maximize takes on a particular functional form. The efficiency and externality terms in the social welfare function are standard---efficiency is total surplus between the worker and the firm ($U_w + U_f$) and the externality is captured by $\gamma U_r$, where $U_r$ is some well-behaved function. The equity term, on the other hand, is less standard. We model the regulator's preference for equality as a quadratic penalty $-\beta (U_w - U_f)^2$, which may appear extreme in its implications. In particular, it suggests that for any positive $\beta,$ the first-best contract results in a perfectly equal distribution of the surplus in all states of the world. 

Although this assumption may not always align with how regulators and lawmakers think about equity in practice, there are two key reasons why it is a useful assumption for understanding the theoretical limits faced by the regulator. 

First, the equality quadratic penalty term is really a stand-in for a broader class of weighted quadratic penalty terms: it captures the idea that in the first-best contract, the regulator wants \emph{a particular} distribution of the surplus. The fact that the regulator wants an equal distribution of the surplus is besides the point---the analysis would be largely unchanged if the SWF featured a weighted penalty term with 
$$-\beta (\alpha U_w - (1-\alpha)U_f)^2$$
for $\alpha \in (0,1).$ For example, with $\alpha = .75,$ the first-best would always feature a 75-25 split of the surplus. 

Second, our focus on a particular distribution of surplus is the most restrictive assumption we could make about the regulator's equality preferences. In practice, lawmakers tolerate inequality up to point. So the first-best distributions of surplus often are not a single division, but instead feature a range of possible ways of dividing the surplus. For example, a lawmaker may only think inequality is objectionable if the worker's utility is less than (greater than) some fraction $\bar{k}$ ($\underline{k}$) of the firm's utility. That is, the regulator's inequity penalty may be 
$$\beta(\ind[\underline{k} < U_w/U_f < \overline{k}]) $$
which implies that the first-best is a set of contracts, rather than a singleton. Since we are interested here in understanding the \emph{limits} on implementation, we focus on singleton case which will be the most limited case. If the first-best featured a range of possible distribution splits, first-best would be ``easier" to implement. In this sense, our assumption about the quadratic penalty term is a limiting case that offers bounds on implementability of more realistic equity preferences.

\subsection{Exogenous income}\label{sec:exogenous income}

The firm and worker may enter the contracting environment with exogenous income. Exogenous income has consequences for the regulator's distributional preferences, and also may affect limited liability constraints, which we have not discussed. With exogenous income, the worker and firm preferences are given by

$$\textbf{Firm: } U_f(q,c; \theta) = y_f+ u_f(q; \theta)-c \hspace{4mm} \textbf{Worker: } U_w(q,c; \theta) = y_w + u_w(q; \theta)+c,$$
where \(y_i\) represents an exogenous and separable component of preferences for agent \(i \in \{w,f\}\). In practice, \(y_w\) may represent workers' wages pinned down by an outside option whereas \(y_f\) is the firm's total revenue net of these wages. 

When there is a regulator with preferences over efficiency and equality, these exogenous parameters will simply effect the water-level of the transfers. An increase (decrease) in \(y_f\) or a decrease (increase) in \(y_w\) will necessitate a higher (lower) transfer in all states. This does not effect the regulator's ability to implement the first-best however, as the first best only depends on the state-dependence components of outcomes. Thus, we have the following proposition. 

\begin{proposition} Assume a regulator is  maximizing social welfare which has efficiency, equity and externality components. Any exogenous and separable components of utility which are state-independent and do not depend on quality, only affect the optimal mechanism by adjusting the default transfer. This does not impact implementability. 

\end{proposition}

\section{Application Derivations and Discussion}\label{sec:App_der}
Assume that the firm can provide additional wage in order to compensate for the risk at the job by paying workers \(c\).
\begin{align*}\label{eq:simplefuncs}
 U_w & = WS(c, q; \theta) = c -(q-\theta)^2  \\
    U_f &= \Pi(c, q) = R - q^2 -c\\
\end{align*}

The regulator wants to maximize 
\[    U_p = WS + \Pi-\beta (WS - \Pi)^2 +\gamma q \]
so that the regulator cares about the firm and worker surplus, the distribution and a term which captures an externality or social value associated with quality.

Summarizing:
\begin{itemize*}
\item \(\theta\): captures the optimal level of safety for the consumer and affects the tradeoff between additional wages versus higher safety
\item R: revenue of the firm net of baseline pay to worker
\item \(c\): compensation
\item q: level of safety measures
\item \(G(\theta)\): Prior over the state. Throughout we will assume \(\theta \sim U(0,1)\)

\end{itemize*}

\subsection{Regulator Problem}

The regulator can choose a default \((q_d, c_d)\) and cannot prevent renegotiation. However, they can impose a minimum, \(\underline{q}\) and maximum quality level, \(\overline{q}\). Without any constraints on the quality level we know that the firm and the consumer will negotiate until \(q_\theta = \frac{\theta}{2}\). This allows us to divvy up the state space into three intervals. We also know that the transfer will be the outcome of the renegotiation from the default and will be given by 
\[c_\theta = c(q_\theta,c_d,q_d;\theta) = c_d +(1-\delta)(-(q_d -\theta)^2+(q_\theta-\theta)^2))-\delta(-q_d^2+q_\theta^2))\]
\[c_\theta = c(q_\theta,c_d,q_d;\theta) = c_d +(1-\delta)(q_\theta^2-q_d^2+2\theta(q_d-q_\theta))-\delta(-q_d^2+q_\theta^2))\]

\[c_\theta = c(q_\theta,c_d,q_d;\theta) = c_d +(1-2\delta)q_\theta^2-(1-2\delta)q_d^2+2(1-\delta)\theta(q_d-q_\theta)\]

\textbf{Binding minimum: } \( \theta< 2\underline{q}\rightarrow q_\theta = \underline{q}\)

\textbf{Unconstrained Interval: } \( 2\underline{q}<\theta< 2\overline{q}\rightarrow q_\theta = \frac{\theta}{2}\)

\textbf{Binding Maximum: } \( \theta> 2\overline{q}\rightarrow q_\theta = \overline{q}\)

\subsubsection{Why Interval Delegation?}\label{sec:AM_application}
Alonso and Matouschek (2008) characterizes conditions under which interval delegation is optimal. Their conditions apply in the case of delegation to a single agent, and have to do with the difference between the agent and the  principal's optimal decisions. 

A key object in their analysis is the \emph{backward bias}, defined as

\[T(\theta) = G(\theta)(q_\theta - E[q^*(z)|z\leq \theta])\]
where \(q^*(z)\) is the regulator's optimal quality choice conditional on bargaining. They show that if \(T''(\theta)>0\) in the relevant range of \(\theta\), then the regulator will choose interval delegation. 

\begin{proposition}[\citet{alonsomatouschek},Proposition 2 (i)]
Let $\Q^*$ be an optimal delegation set. Then if $T(\theta)$ is strictly convex then $Q^*$ contains either no decision, one decision, or an interval of decisions.
\end{proposition} 

We can apply this result in our example to confirm that the regulator would choose interval delegation. In our example, the backward bias is
\[T(\theta) = \theta \left(\frac{\theta}{2} - \frac{1}{\theta}\int_0^\theta (q^*(z)) dz\right) = \frac{\theta^2}{2}-\int_0^\theta q^*(z)dz\]
and its second derivative is
\begin{equation}\label{eq:AM_convexbias}
    T''(\theta) = 1-\frac{\partial q^*(\theta)}{\partial \theta}.
\end{equation}
Next we need to solve for $q^*(\theta)$, the regulator's optimal quality outcome conditional on the state. Taking the default as given, $q^*(\theta)$ is defined by the problem 
\begin{equation}\label{eq:AM_qopt}
    q^*(\theta) = \argmax_q \left[R -(q-\theta)^2-q^2 - \beta(-(q-\theta)^2+q^2 + 2c )^2+\gamma q\right].
\end{equation}
The term $c$ in the maximand above is in fact a function of $q, \delta,$ and $q_d.$ In this set up, we are treating $\delta$ and $q_d$ as exogenous, and can rewrite $c$ explicitly, 
\[c =c_d +(1-2\delta)q^2-(1-2\delta)q_d^2+2(1-\delta)\theta(q_d-q). \]
 Now plugging this expression for $c$ into \eqref{eq:AM_qopt} and taking the first order condition, we get
 \begin{equation}\label{eq:AM_foc}
 -4q^*(\theta)+2\theta+\gamma -2\beta\left(-R+2c_\theta-(q^*(\theta)-\theta)^2+q^*(\theta)^2\right) \left(2\theta+4\left[(1-2\delta)q^*(\theta)-\theta(1-\delta)\right]\right)=0. 
  \end{equation}
To make this expression concrete, consider the simplest case when \(\delta = .5\). In this case the first order condition \eqref{eq:AM_foc} simplifies to 
\[ q^*(\theta)=\frac{2\theta+\gamma}{4}.\]
That is, the regulator's optimal $q$ is the agent-optimal $q$ shifted by $\frac{\gamma}{4}$. Substituting $q^*(\theta)$ into the equation for the second derivative of the backward bias \eqref{eq:AM_convexbias} yields
\[T''(\theta) = \frac{1}{2}.\]
By proposition 2 in Alonso and Matouschek (2008), the fact that $T''(\theta)>0$ implies that the regulator's optimal delegation set is either no decision, one decision or an interval.

Next consider the case where \(\delta = 0\). In this case, the regulator's first-order condition in \eqref{eq:AM_foc} simplifies to
\[-4q^*(\theta)+2\theta+\gamma -2\beta\left(-R+2c_\theta + 2q^*(\theta)\theta -\theta^2\right) \left(4q^*(\theta)-2\theta\right)=0.\]
Furthermore, note that the default transfer will always cancel the initial inequality such that we can consider \(R=0\) without loss of generality. Substituting in the transfer we get
\[-4q^*(\theta)+2\theta+\gamma -2\beta\left(2\left(c_d +q^*(\theta)^2-q_d^2+2\theta(q_d-q^*(\theta))\right) + 2q^*(\theta)\theta -\theta^2\right) \left(4q^*(\theta)-2\theta\right)=0\]
Implicitly differentiating yields: 
\begin{align*}
    -4\frac{\partial q^*(\theta)}{\partial \theta} + 2 -2\beta\left(4\frac{\partial q^*(\theta)}{\partial \theta}-2\right)(2c_\theta + 2q^*(\theta)\theta -\theta^2)\\
    -2\beta(4q^*(\theta)-2\theta)\left(4q^*(\theta)\frac{\partial q^*(\theta)}{\partial \theta}+2(q_d-q^*(\theta))-2\theta\frac{\partial q^*(\theta)}{\partial \theta}\right)
\end{align*}
Simplifying, yields
\[\frac{\partial q^*(\theta)}{\partial \theta} = \frac{1+2\beta(2c_\theta -\theta^2+2\theta q_d-4q^*(\theta)(q_d-q^*(\theta)))}{2+4\beta(2c_\theta + 2q^*(\theta)\theta -\theta^2+(2q^*(\theta)-\theta)^2)}\]
As long as this expression is less than 1, the second derivative of the backward bias in \eqref{eq:AM_convexbias} will be positive. We can note a few circumstances under which this expression is less than one. First, when the equity parameter \(\beta\) is small, this expression is less than 1. Note that this also will be the case when the changes in equity are relatively small from figure \ref{fig:opt} that the impact on inequity is lower in the middle than at the extremes which will make this condition more likely to hold. 

A similar analysis can be undertaken for the case of \(\delta =1\).

\subsubsection{Solving the Regulator's Problem}
The regulator problem is then given by 
\begin{align*}
    \max_{\{(c_d,q_d,\underline{q},\overline{q})} &\int_{\underline{\theta}}^{2\underline{q}} \left(R-(\underline{q}-\theta)^2-\underline{q}^2- \beta\left(-R+2c(\underline{q},c_d,q_d;\theta)-(\underline{q}-\theta)^2+\underline{q}^2\right)^2+\gamma \underline{q}\right)dG(\theta) + \\
   &\int_{2\underline{q}}^{2\overline{q}} \left(R -\frac{\theta^2}{2} - \beta\left(-R+2c(\theta/2,c_d,q_d;\theta)\right)^2+\gamma \frac{\theta}{2}\right)dG(\theta) + \\ 
   & \int_{2\overline{q}}^{\overline{\theta}} \left(R-(\overline{q}-\theta)^2-\overline{q}^2- \beta\left(-R+2c(\overline{q},c_d,q_d;\theta)-(\overline{q}-\theta)^2+\overline{q}^2\right)^2+\gamma \overline{q}\right)dG(\theta)
\end{align*}

Subject to the constraints:
\begin{align*}
    2\underline{q} &\geq \underline{\theta} & & \\
    2\overline{q} &\leq \overline{\theta} & & \\
    2\overline{q} &\geq \underline{\theta} & & \\
    \underline{q} &\leq q_d \leq \overline{q} & & \\
\end{align*}
It is useful here to totally differentiate \(c(q_\theta,c_d,q_d;\theta)\)

\[dc_\theta = dc_d + 2\left[(2\delta-1)q_d+\theta(1-\delta)\right]dq_d + 2\left[(1-2\delta)q_\theta-\theta(1-\delta)\right]dq_\theta \]

Now we can take first order conditions. 

\[\frac{\partial V}{\partial \underline{q}} = \int_{\underline{\theta}}^{2\underline{q}} \left( -4\underline{q}+2\theta+\gamma -2\beta\left(-R+2c_\theta-(\underline{q}-\theta)^2+\underline{q}^2\right) \left(2\theta+4\left[(1-2\delta)\underline{q}-\theta(1-\delta)\right]\right)\right)dG(\theta)\]

\[\frac{\partial V}{\partial \overline{q}} = \int_{2\overline{q}}^{\overline{\theta}}\left( -4\overline{q}+2\theta+\gamma -2\beta\left(-R+2c_\theta-(\overline{q}-\theta)^2+\overline{q}^2\right) \left(2\theta+4\left[(1-2\delta)\overline{q}-\theta(1-\delta)\right]\right)\right)dG(\theta)\]

\begin{align*}
    \frac{\partial V}{\partial c_d} = 0 & = \frac{-R}{2} +\\ &\int_{\underline{\theta}}^{2\underline{q}} \left(c_d +(1-2\delta)\underline{q}^2-(1-2\delta)q_d^2+2(1-\delta)\theta(q_d-\underline{q})-\frac{\theta^2}{2}+\underline{q}\theta\right)dG(\theta) + \\
   &\int_{2\underline{q}}^{2\overline{q}} \left(c_d +(1-2\delta)\frac{\theta^2}{4}-(1-2\delta)q_d^2+2(1-\delta)\theta\left(q_d-\frac{\theta}{2}\right)\right)dG(\theta) + \\ 
   & \int_{2\overline{q}}^{\overline{\theta}} \left(c_d +(1-2\delta)\overline{q}^2-(1-2\delta)q_d^2+2(1-\delta)\theta(q_d-\overline{q}))-\frac{\theta^2}{2}+\overline{q}\theta\right)dG(\theta)
\end{align*}

\begin{align*}
    \frac{\partial V}{\partial q_d} = 0 =  &\int_{\underline{\theta}}^{2\underline{q}} \left(-R+2c_\theta-(\underline{q}-\theta)^2+\underline{q}^2\right)\left[(2\delta-1)q_d+\theta(1-\delta)\right]dG(\theta) + \\
   &\int_{2\underline{q}}^{2\overline{q}} \left(-R+2c_\theta\right)\left[(2\delta-1)q_d+\theta(1-\delta)\right]dG(\theta) + \\ 
   & \int_{2\overline{q}}^{\overline{\theta}} \left(-R+2c_\theta-(\overline{q}-\theta)^2+\overline{q}^2\right)\left[(2\delta-1)q_d+\theta(1-\delta)\right]dG(\theta)
\end{align*}

\subsection{Equal Bargaining $\delta = .5$}

It is useful to start by considering the case where the firm and worker have equal bargaining positions. In this case they will split any surplus generated from renegotiation equally. Importantly, the regulator's equality term aims to equate worker surplus and firm profits. Thus, the bargaining conditions are aligned with the regulator's incentives as far as equity is concerned. 

We also know that the regulator's externality term favors high quality. Combined this means that the regulator would never want to implement a maximum quality level. However, it would implement a minimum quality level. When \(\delta = .5\) the FOC's simplify to

\[\frac{\partial V}{\partial \underline{q}} = \int_{\underline{\theta}}^{2\underline{q}} \left( -4\underline{q}+2\theta+\gamma \right)dG(\theta)\]

\[\frac{\partial V}{\partial \overline{q}} = \int_{2\overline{q}}^{\overline{\theta}}\left( -4\overline{q}+2\theta+\gamma \right)dG(\theta)\]

\[
    \frac{\partial V}{\partial c_d} = 0  = \frac{-R}{2} + \int_{\underline{\theta}}^{\overline{\theta}} \left(c_d +\theta q_d-\frac{\theta^2}{2}\right)dG(\theta) 
\]

\begin{align*}
    \frac{\partial V}{\partial q_d} = 0 = \int_{\underline{\theta}}^{\overline{\theta}} \theta\left(-R+2\left(c_d +\theta\left(q_d-\frac{\theta}{2}\right)\right)\right)dG(\theta)
\end{align*}

Note that as expected there is no dependence on \(\beta\). We can also solve explicitly the regulator's problem: 

\[\underline{q} = \frac{\gamma}{2}\]

\[\overline{q} = \frac{\overline{\theta}}{2}= \frac12 \]
\[c_d = \frac{R}{2} -  \mu_{\theta}q_d + \frac{\mu_{\theta^2}}{2}\]

\[q_d = \frac{1}{2} \frac{\mu_{\theta^3}-\mu_{\theta^2}\mu_\theta}{\mu_{\theta^2}-\mu_{\theta}\mu_\theta}\]

Note that the term inside the integral for \(\frac{\partial V}{\partial \overline{q}}\) is always positive and thus, the maximum is at a corner solution and thus, does not bind. We also see that \(\underline{q}\) is increasing in \(\gamma\) as that is the only way to ensure a higher quality level when there is renegotiation. Lastly we have that the default transfer with our assumed distribution is \(c_d = \frac{R}{2} -\frac{1}{12}\) and \(q_d = \frac{1}{2}\). To interpret this we note that the default transfer's first job is to equate the initial surplus \(R\) between the two parties. Then we see that the quality default is the expected optimal for the worker. Thus, the regulator is paying the worker who has state dependent preferences in terms of the quality whereas the firm is being paid in terms of a negative default transfer beyond the initial redistribution. Interestingly the presence of the minimum does not affect the default quality and transfer. This is because the only role of the defaults are to ensure equality. Since the two parties always equally divide the surplus, the realized outcomes do not affect equality, only the relationship between the default and the state. 

It is interesting to consider an alternative distribution. Below are the results when \(g(\theta) = 1+2(\theta-.5) \quad \text{for} \quad \theta \in \left(0,1\right)\). With this distribution we will now get a higher minimum quality level and a different default quality and transfer. Specifically, 

\[\underline{q} = \frac{3\gamma }{4}\]
\[\overline{q} = \frac12\]
\[c_d = \frac{R}{2} -  \mu_{\theta}q_d + \frac{\mu_{\theta^2}}{2} = \frac{R}{2} -  \frac{3}{20}\]
\[q_d = \frac{1}{2} \frac{\mu_{\theta^3}-\mu_{\theta^2}\mu_\theta}{\mu_{\theta^2}-\mu_{\theta}\mu_\theta} = \frac{3}{5}\]

Note that the default quality has gone up, the default transfer has gone down and the responsiveness to externalities has gone up. All of these make sense given that the benefit of a high default is higher now that the typical state is higher. Furthermore, we can compare the default quality to the expected worker optimal as before. The expected value of the state is now \(2/3\), which means that although the mean value increasing has pushed up the default quality, the skewed distribution has made it so that the default quality is no longer as high as the workers expected optimal. 

\subsection{Firm Power $\delta = 0$}

The most interesting case and the case that gives us intuition for the realm where we may expect the regulator to be operating is when the firm has all of the bargaining power. From our discussion in implementation theory, we know that this is the state where the regulator is in the best position to implement something approximating first best because the worker is more sensitive to the true state.

Substituting \(\delta=0\) into the FOCs yields

\[\frac{\partial V}{\partial \underline{q}} = \int_{\underline{\theta}}^{2\underline{q}} \left( -4\underline{q}+2\theta+\gamma -2\beta\left(-R+2c_\theta-(\underline{q}-\theta)^2+\underline{q}^2\right) \left(4\underline{q}-2\theta\right)\right)dG(\theta)\]

\[\frac{\partial V}{\partial \overline{q}} = \int_{2\overline{q}}^{\overline{\theta}}\left( -4\overline{q}+2\theta+\gamma -2\beta\left(-R+2c_\theta-(\overline{q}-\theta)^2+\overline{q}^2\right) \left(4\overline{q}-2\theta\right)\right)dG(\theta)\]

For the solver it is useful to rewrite these equations: 

\begin{align*}
    \frac{\partial V}{\partial \underline{q}} = 
    (2\underline{q})\left(\gamma -2\underline{q}-4\beta\underline{q}\left(2c_d+\frac{8}{3}q_d\underline{q}-2q_d^2-R\right)\right) =0 
\end{align*}

First thing to note is that for weakly positive \(\underline{q}\) and \(\gamma =0\), this equation holds when \(\beta = 2\). You can also rewrite it defining \(\underline{d} = 2\underline{q}\). 

\[   \frac{\partial V}{\partial \underline{q}} = \underline{d}\left(\gamma -\underline{d}-2\beta\underline{d}\left(2c_d+\frac{4}{3}q_d\underline{d}-2q_d^2-R\right)\right) =0 \]

\[\frac{\partial V}{\partial \overline{q}}=0=(1-2\overline{q})(1+\gamma-2\overline{q}+\frac{\beta}{3}(1-2\overline{q})(-3+12c_d+4(4-3q_d)q_d+4\overline{q}(4q_d-3)-6R))\]
Similarly, defining \(\overline{d}=1-2\overline{q}\)
\[\frac{\partial V}{\partial \overline{q}}=0=\overline{d}\left(\gamma+\overline{d}+2\beta\overline{d}\left(2c_d-\frac{4}{3}q_d\overline{d}-2q_d^2+4q_d-\frac{3}{2}-R\right)\right)\]
We can solve these two explicitly in the case wehere \(\gamma =0\) and \(\overline{d}\) and \(\underline{d}\) are weakly positive.
\[\underline{d} = \frac{3}{4q_d}\left(\frac{-1}{2\beta}-2c_d+2q_d^2 -R\right)\]
\[\overline{d} = \frac{\left(\frac{-1}{2\beta}-2c_d+2q_d^2-R +4q_d-\frac{3}{2}\right)}{\frac{4}{3}q_d-1}\]

We can also substitute in for \(c_d=\frac{R}{2} + \frac{3\mu_{\theta^2}}{4} +q_d^2 - 2q_d \mu_\theta\).

\[\underline{d} = \frac{\frac{-1}{2\beta}-\frac{3\mu_{\theta^2}}{2} +4q_d \mu_\theta }{\frac{4}{3}q_d}\]

\[\overline{d} = \frac{\left(\frac{-1}{2\beta}-\frac{3\mu_{\theta^2}}{2}  +4q_d (1+\mu_\theta)-\frac{3}{2}\right)}{\frac{4}{3}q_d-1}\]

We can then differentiate these with respect to \(q_d\)

\[d\underline{d} = \frac{\frac{16}{3}q_d\mu_\theta-\frac{4}{3}\left(\frac{-1}{2\beta}-\frac{3\mu_{\theta^2}}{2} +4q_d \mu_\theta \right)}{(4q_d/3)^2}dq_d\]
\[d\overline{d} = \frac{4(q_d+\mu_\theta)(\frac{4}{3}q_d-1)-\frac{4}{3}\left(\frac{-1}{2\beta}-\frac{3\mu_{\theta^2}}{2}  +4q_d (1+\mu_\theta)-\frac{3}{2}\right)}{(\frac{4}{3}q_d-1)^2}dq_d\]
At \(\beta = 2\), \(q_d = 3/8\). These two equations simplify to 

\[d\underline{d} = 4 dq_d\]
\[d\overline{d} = -7 dq_d\]

\begin{align*}
    \frac{\partial V}{\partial c_d} = 0 & = \frac{-R}{2} +\\ &\int_{\underline{\theta}}^{2\underline{q}} \left(c_d +\underline{q}^2-q_d^2+2\theta(q_d-\underline{q})-\frac{\theta^2}{2}+\underline{q}\theta\right)dG(\theta) + \\
   &\int_{2\underline{q}}^{2\overline{q}} \left(c_d +\frac{\theta^2}{4}-q_d^2+2\theta\left(q_d-\frac{\theta}{2}\right)\right)dG(\theta) + \\ 
   & \int_{2\overline{q}}^{\overline{\theta}} \left(c_d +\overline{q}^2-q_d^2+2\theta(q_d-\overline{q})-\frac{\theta^2}{2}+\overline{q}\theta\right)dG(\theta)
\end{align*}

Or 
\[\frac{\partial V}{\partial c_d} = 0=\frac{-R}{2}+c_d-\frac{1}{6}+\frac{1}{6}\left(6(1-q_d)q_d+\overline{q}(-3+6\overline{q}-4\overline{q}^2)+4q_l^3\right)\]

\begin{align*}
    \frac{\partial V}{\partial q_d} = 0 =  &\int_{\underline{\theta}}^{2\underline{q}} \left(-R+2c_\theta-(\underline{q}-\theta)^2+\underline{q}^2\right)\left[-q_d+\theta\right]dG(\theta) + \\
   &\int_{2\underline{q}}^{2\overline{q}} \left(-R+2c_\theta\right)\left[-q_d+\theta\right]dG(\theta) + \\ 
   & \int_{2\overline{q}}^{\overline{\theta}} \left(-R+2c_\theta-(\overline{q}-\theta)^2+\overline{q}^2\right)\left[-q_d+\theta\right]dG(\theta)
\end{align*}

or 
\[\frac{\partial V}{\partial q_d} = 0 =-3+\left(c_d+\frac{-R}{2}\right)(12-24q_d)-36q_d^2+24q_d^3-4 \overline{q}(2-3\overline{q}+2\overline{q}^3)+8\underline{q}^4+4q_d(5+\overline{q}(3-6\overline{q}+4\overline{q}^2)-4\underline{q}^3)\]

Furthermore, we can substitute in the value for \(c_d\) to get: 

\[q_d=\frac{1+2\overline{q}-8(1-\overline{q})\overline{q}^3+8(1-\underline{q})\underline{q}^3}{4}\]

\[4dq_d=\left(2-8(3-4\overline{q})\overline{q}^2 \right)d\overline{q}+8\left(3-4\underline{q}\right)\underline{q}^2 d\underline{q}\]

As before it is useful to solve these FOCs for when there is neither a minimum nor maximum. For instance, in the limit as \(\gamma \rightarrow 0\) and \(\beta < 2\). In this case, 

\[c_d = \frac{R}{2} + \frac{3\mu_{\theta^2}}{4} +q_d^2 - 2q_d \mu_\theta = \frac{R}{2} +\frac{1}{64}\]

\[q_d = \frac{3}{8}\frac{\mu_{\theta^3}-\mu_{\theta^2}\mu_{\theta}}{\mu_{\theta^2}-\mu_{\theta}\mu_{\theta}}=\frac{3}{8}\]

Note that the default quality is lower when the workers have a worse bargaining position. This because any renegotiation away from the default will result in higher surplus for the firm relative to the worker. Thus, in order to reduce inequality across states it is useful to reduce the extent of renegotiation and adjust the transfer. 

Similarly, we can solve for \(\underline{q}\) when \(\beta \rightarrow 0\) resulting in \(\underline{q} = \frac{\gamma}{2}\), which unsurprisingly is the same as in the equal bargaining case. As in either case the minimum is defined by efficiency and externality concerns and are independent of equality. 

It is also useful to consider how the minimum adjusts with inequality. In order to gain some intuition consider we can evaluate \(\frac{\partial^2 V}{\partial \underline{q}\partial \beta}\) at \(\beta = 0\)

\[\frac{\partial^2 V}{\partial \underline{q}\partial \beta}|_{\beta=0} =  
    -4\underline{q}\left(2c_d+\frac{8}{3}q_d\underline{q}-2q_d^2-R\right) \]

We can substitute in the values for \(c_d\) and \(q_d\) to get 

\[\frac{\partial^2 V}{\partial \underline{q}\partial \beta}|_{\beta=0} = -4\underline{q}\left(\frac{3}{2}\mu_{\theta^2}-4q_d\mu_{\theta}+\frac{8}{3}q_d\underline{q}\right) = -4\underline{q}\left(\frac{1}{2}-\frac{3}{4}+\underline{q}\right)\]

This means that starting from \(\underline{q}=0\), the minimum is increasing in \(\beta\). In general we expect that the interval over which the the firm and worker are able to choose quality is decreasing in concerns over equality. Note that with the firm making take it or leave it offers to the worker,

\subsection{Worker Power $\delta = 1$}
Another special case is when the workers have all of the bargaining power. The difficulty here from the standpoint of the regulator is that the firm does not have state dependent preferences. This highlights the importance of state dependence as highlighted in the implementation discussion above. Thus, the default which is used to smooth the necessary transfer across states can no longer perform this function. We can see this directly in the first order conditions by plugging in \(\delta = 1\). 

\[c_\theta = c(q_\theta,c_d,q_d;\theta) = c_d -q_\theta^2+q_d^2\]

Note that the transfer does not depend on the state except through the level of quality. 

\[\frac{\partial V}{\partial \underline{q}} = \int_{\underline{\theta}}^{2\underline{q}} \left( -4\underline{q}+2\theta+\gamma -2\beta\left(-R+2c_\theta-(\underline{q}-\theta)^2+\underline{q}^2\right) \left(-2\underline{q}\right)\right)dG(\theta)\]

\[\frac{\partial V}{\partial \overline{q}} = \int_{2\overline{q}}^{\overline{\theta}}\left( -4\overline{q}+2\theta+\gamma -2\beta\left(-R+2c_\theta-(\overline{q}-\theta)^2+\overline{q}^2\right) \left(-2\overline{q}\right)\right)dG(\theta)\]

\begin{align*}
    \frac{\partial V}{\partial c_d} = 0 & = \frac{-R}{2} +\\ &\int_{\underline{\theta}}^{2\underline{q}} \left(c_d -\underline{q}^2+q_d^2-\frac{\theta^2}{2}+\underline{q}\theta\right)dG(\theta) + \\
   &\int_{2\underline{q}}^{2\overline{q}} \left(c_d -\frac{\theta^2}{4}+q_d^2\right)dG(\theta) + \\ 
   & \int_{2\overline{q}}^{\overline{\theta}} \left(c_d -\overline{q}^2+q_d^2-\frac{\theta^2}{2}+\overline{q}\theta\right)dG(\theta)
\end{align*}

\begin{align*}
    \frac{\partial V}{\partial q_d} = 0 =  &\int_{\underline{\theta}}^{2\underline{q}} \left(-R+2\left(c_d -\underline{q}^2+q_d^2\right)-\theta^2 +2\underline{q}\theta\right)\left[-q_d\right]dG(\theta) + \\
   &\int_{2\underline{q}}^{2\overline{q}} \left(-R+2\left(c_d -\frac{\theta^2}{4}+q_d^2\right)\right)\left[-q_d\right]dG(\theta) + \\ 
   & \int_{2\overline{q}}^{\overline{\theta}} \left(-R+2\left(c_d -\overline{q}^2+q_d^2\right)-\theta^2 +2\overline{q}\theta\right)\left[-q_d\right]dG(\theta)
\end{align*}
Note that because \(q_d\) is just a constant we can eliminate it from the last expression which shows that when the workers have all of the bargaining power, the transfer and quality level are only jointly determined. We can show this by rewriting the last equation

\begin{align*}
    \frac{\partial V}{\partial q_d} = 0 = \frac{-R}{2}+ &\int_{\underline{\theta}}^{2\underline{q}} \left(c_d -\underline{q}^2+q_d^2-\frac{\theta^2}{2} +\underline{q}\theta\right)dG(\theta) + \\
   &\int_{2\underline{q}}^{2\overline{q}} \left(c_d -\frac{\theta^2}{4}+q_d^2\right)dG(\theta) + \\ 
   & \int_{2\overline{q}}^{\overline{\theta}} \left(c_d -\overline{q}^2+q_d^2-\frac{\theta^2}{2} +\overline{q}\theta\right)dG(\theta)
\end{align*}
which is the same as the first order condition for \(c_d\). If the maximum and minimum are not binding this simply implies that \(\frac{R}{2} -c_d - q_d^2 = \frac{\mu_{\theta^2}}{4} =-\frac{1}{12}\), which is the expected losses that the workers will experience from the optimal level of quality. Thus, the firm will always receive \(\frac{R+w}{2}- \frac{1}{12}\) as will the workers in expectation. If the minimum is binding, the workers will be made better off relative to the firm in low states of the world and thus, the default will become more favorable to the firm. Similarly, if there is a maximum, then the maximum will make the firm relatively better off compared to the workers and thus, the default will be less favorable to the firm. 

Furthermore, we can define the default value of the firm as \(F_d =  -c_d - q_d^2\) and reconsider the regulator problem as choosing \(\underline{q},\overline{q},F_d\). Rewriting and simplifying the FOCs:

\[\frac{\partial V}{\partial \underline{q}} = \int_{\underline{\theta}}^{2\underline{q}} \left( -4\underline{q}+2\theta+\gamma +4\underline{q}\beta\left(-R+2\left(-F_d -\underline{q}^2\right)+2\underline{q}\theta-\theta^2\right)\right)dG(\theta)\]

\[\frac{\partial V}{\partial \overline{q}} = \int_{2\overline{q}}^{\overline{\theta}}\left( -4\overline{q}+2\theta+\gamma +4\overline{q}\beta\left(-R+2\left(-F_d -\overline{q}^2\right)-(\overline{q}-\theta)^2+\overline{q}^2\right)\right)dG(\theta)\]

\begin{align*}
    \frac{\partial V}{\partial F_d} = 0 & = \frac{-R}{2} +\\ &\int_{\underline{\theta}}^{2\underline{q}} \left(-F_d  -\underline{q}^2-\frac{\theta^2}{2}+\underline{q}\theta\right)dG(\theta) + \\
   &\int_{2\underline{q}}^{2\overline{q}} \left(-F_d -\frac{\theta^2}{4}\right)dG(\theta) + \\ 
   & \int_{2\overline{q}}^{\overline{\theta}} \left(-F_d  -\overline{q}^2-\frac{\theta^2}{2}+\overline{q}\theta\right)dG(\theta)
\end{align*}

Solving for \(F_d\) yields:

\[
F_d = \frac{-R}{2}-\frac{4}{3}\underline{q}^3-\frac{2}{3}(\overline{q}^3-\underline{q}^3)+\frac{4}{3}\overline{q}^3-\overline{q}^2-\frac{1}{6}+\overline{q}/2\]

\[
F_d = \frac{-R}{2}-\frac{2}{3}\underline{q}^3+\frac{2}{3}\overline{q}^3-\overline{q}^2-\frac{1}{6}+\overline{q}/2\]

It is important here to note that \(\frac{\partial F_d }{\partial \underline{q}} < 0\) and \(\frac{\partial F_d }{\partial \overline{q}} > 0\). These signs are intuitive when we consider how the worker's utility changes in the range where either the minimum or maximum is binding. In either case, the worker's utility is decreasing the tighter is the constraint because they get all of the surplus from renegotiation. Thus,  when the minimum \(\underline{q}\) raises the worker's position deteriorates which induces an optimal reduction in the firm surplus. Similarly when \(\overline{q}\) decreases, and thus, the constraint binds more tightly, the firm value \(F_d\) must decrease. 

However, this immediately suggests that the maximum must never bind because we know that the worker's utility is always below the firm's in the high state. In the low state \(\theta =0\), the worker receives its optimal quality as well as receiving the the maximum transfer from the firm. In the highest state, the firm is made the worst off from renegotiation which minimizes the value the workers receive from renegotiation. In this case the maximum can only increase inequality.

In order to see this we can look directly at the inequality term. We know that the worker's utility is given by \( \frac{R}{2} -(q_\theta-\theta)^2 -F_d-q_\theta^2  \)
Meanwhile the firm receives \(\frac{R}{2}+F_d\). The difference then is given by 

\[Eq = W - F = -(q_\theta-\theta)^2 -2F_d-q_\theta^2 \]
With renegotiation, we can plug in \(\theta/2\) and get

\[Eq = -\frac{\theta^2}{2} -2F_d \]

This is a decreasing function of the state in the relevant range \(\theta>0\). Thus, in order to minimize the deviation it must be that \(F_d > -\frac{\overline{\theta}^2}{4}\), so that \(Eq < 0\) when \(\theta = \overline{\theta}\). Now consider decreasing \(q_\theta\) where \(\theta = \overline{\theta}\). In this case, 

\[\frac{\partial Eq }{\partial q_\theta} = -2(q_\theta-\theta)-2q_\theta\]
Note that this is positive when \(q_\theta < \frac{\theta}{2}\). Thus, any decrease in the maximum will lead to a decrease in \(Eq\) which given that \(Eq\) is negative implies greater inequality. Thus, we know that \(\overline{q} = .5\) in this setting.

Now we can solve for the minimum by substituting in \(F_d = \frac{-R}{2}-\frac{2}{3}\underline{q}^3-\frac{1}{12}\).

\[\frac{\partial V}{\partial \underline{q}} = \int_{\underline{\theta}}^{2\underline{q}} \left( \underbrace{-4\underline{q}+2\theta}_{\textbf{efficiency}}+\underbrace{\gamma}_{\textbf{externality}} +\underbrace{4\underline{q}\beta\left(\frac{4}{3}\underline{q}^3+\frac{1}{6} -2\underline{q}^2+2\underline{q}\theta-\theta^2\right)}_{\textbf{equality}}\right)dG(\theta)\]
Note here that it is clear that losses due to efficiency will reduce the value of the minimum, the externality will push the minimum up and the equality term will also lead to greater minimums.

\bibliography{refs.bib}

\pagebreak

\end{document}